\documentclass[10pt,sigconf,letterpaper]{acmart}
\usepackage[english]{babel}
\usepackage{wasysym}
\usepackage{subcaption}
\usepackage{tabularx}
\usepackage{tabulary}
\usepackage{booktabs}
\usepackage{colortbl}
\usepackage{multirow}
\usepackage{color}
\usepackage{soul}
\usepackage{chngpage}
\usepackage{caption}
\usepackage{cleveref}
\definecolor{lightgray}{gray}{0.95}

\usepackage{listings}
\usepackage{xcolor}
\usepackage{enumitem}

\crefformat{section}{\S#2#1#3}
\definecolor{pastelred}{RGB}{255,153,153}
\definecolor{pastelblue}{RGB}{204,255,255}
\definecolor{pastelyellow}{RGB}{255,255,204}
\DeclareRobustCommand{\hlred}[1]{{\sethlcolor{pastelred}\hl{#1}}}
\DeclareRobustCommand{\hlblue}[1]{{\sethlcolor{pastelblue}\hl{#1}}}
\DeclareRobustCommand{\hlyellow}[1]{{\sethlcolor{pastelyellow}\hl{#1}}}

\colorlet{punct}{red!60!black}
\definecolor{background}{HTML}{EEEEEE}
\definecolor{delim}{RGB}{20,105,176}
\colorlet{numb}{magenta!60!black}
\definecolor{olivegreen}{HTML}{556b2f}
\definecolor{Gray}{gray}{0.9}

\lstdefinelanguage{json}{
  basicstyle=\ttfamily\scriptsize,
    numbers=left,
    numberstyle=\scriptsize,
    stepnumber=1,
    numbersep=8pt,
    showstringspaces=false,
    breaklines=true,
    frame=lines,
    backgroundcolor=\color{background},
    literate=
     *{0}{{{\color{numb}0}}}{1}
      {1}{{{\color{numb}1}}}{1}
      {2}{{{\color{numb}2}}}{1}
      {3}{{{\color{numb}3}}}{1}
      {4}{{{\color{numb}4}}}{1}
      {5}{{{\color{numb}5}}}{1}
      {6}{{{\color{numb}6}}}{1}
      {7}{{{\color{numb}7}}}{1}
      {8}{{{\color{numb}8}}}{1}
      {9}{{{\color{numb}9}}}{1}
      {:}{{{\color{punct}{:}}}}{1}
      {,}{{{\color{punct}{,}}}}{1}
      {\{}{{{\color{delim}{\{}}}}{1}
      {\}}{{{\color{delim}{\}}}}}{1}
      {[}{{{\color{delim}{[}}}}{1}
      {]}{{{\color{delim}{]}}}}{1},
}

\makeatletter
\def\subsubsection{\@startsection{subsubsection}{3}%
  \z@{.5\linespacing\@plus.7\linespacing}{.1\linespacing}%
  {\normalfont\itshape}}
\makeatother

\renewcommand{\paragraph}[1]{\vspace*{0.03in}\noindent\textbf{#1}}

\usepackage{microtype}
\usepackage[font=small,format=plain,labelfont=bf,textfont=it]{caption}

\newcommand{\todo}[1]{\textcolor{red}{#1}}
\renewcommand{\todo}[1]{\textcolor{red}{}}

\newcommand{\eg}{{\it e.g.}}
\newcommand{\ie}{{\it i.e.}}
\newcommand{\etal}{{\it et al.}}
\newcommand{\sysname}{Traffic Refinery}
\newcommand{\system}{Traffic Refinery}

\title{\system: Cost-Aware Data Representation for Machine Learning on Network Traffic}

\author{
  Francesco Bronzino$^1$$^\ast$, 
  Paul Schmitt$^2$$^\ast$, 
  Sara Ayoubi$^3$,
} 
\thanks{$^\ast$Co-first authors}
\author{
  Hyojoon Kim$^2$, 
  Renata Teixeira$^4$,
  Nick Feamster$^5$}

\affiliation{
  \institution{
    \hspace{0mm}$^1$Universit\'{e} Savoie Mont Blanc \hspace{5mm}$^2$Princeton University}
}
\affiliation{
  \institution{
    \hspace{0mm}$^3$Nokia Bell Labs
    \hspace{5mm}$^4$Inria
    \hspace{5mm}$^5$University of Chicago}
}

\begin{document}

\setcopyright{none}
\settopmatter{printacmref=false} 
\renewcommand\footnotetextcopyrightpermission[1]{} 
\pagestyle{plain} 

\begin{sloppypar}

\begin{abstract}
Network management often relies on machine learning to make predictions
about performance and security from network traffic.
Often, the representation of the traffic is as important as the choice
of the model. The features that the model relies on, and the
representation of those features, 
ultimately determine model accuracy, as well as where and whether
the model can be deployed in practice. 
Thus, the design and evaluation of these models ultimately requires 
    understanding not only model accuracy but also the systems
    costs associated with deploying the model in an operational network.
Towards this goal, this paper develops a new framework and system that enables a joint
evaluation of both the conventional notions of machine learning performance
(\eg, model accuracy) and the systems-level costs of different representations
of network traffic. We highlight these two dimensions for two practical network
management tasks, video streaming quality inference and malware detection, to
demonstrate the importance of exploring different representations to find the
appropriate operating point. We demonstrate the benefit of exploring a range of
representations of network traffic and present \system{}, a proof-of-concept
implementation that both monitors network traffic at 10~Gbps and
transforms traffic in real time to produce a variety of feature
representations for machine learning. \system{} both highlights this
design space and makes it possible to explore
different representations for learning, balancing systems costs related to
feature extraction and model training against model accuracy.
\end{abstract}

\maketitle

\section{Introduction}

\begin{figure}[t]
\begin{center}
    \includegraphics[width=\linewidth]{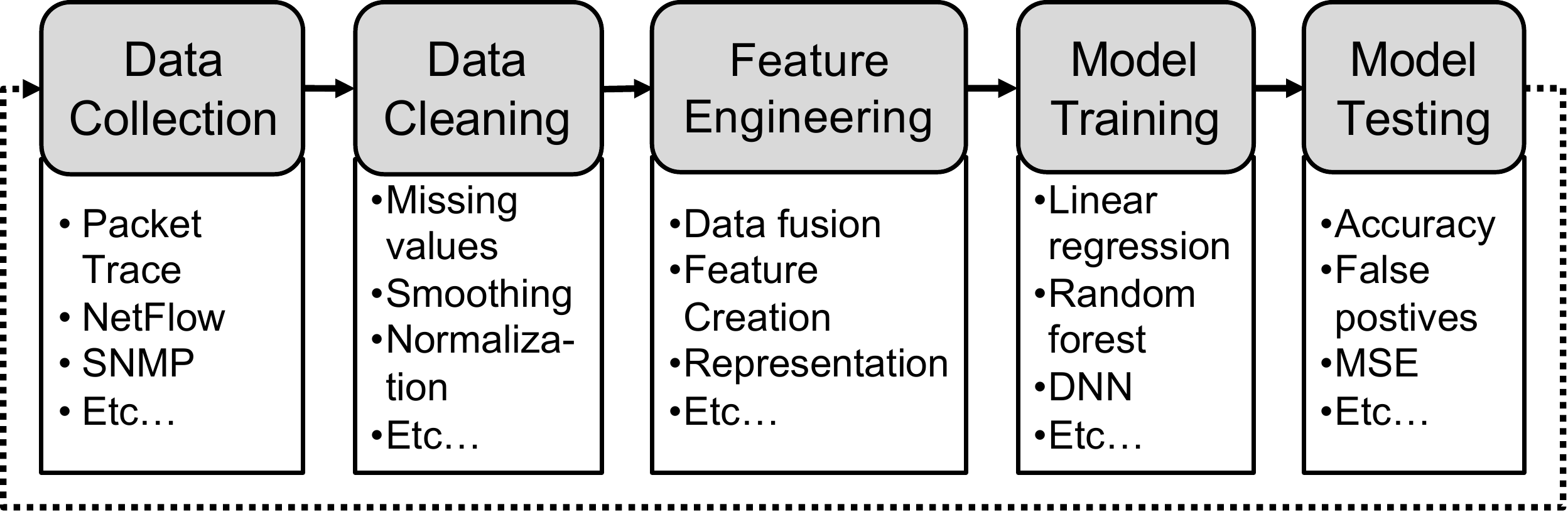}
    \caption{Typical pipeline for model design in network inference.}
    \label{fig:traditional_cycle}
\end{center}
\end{figure}

Network management tasks commonly rely on the ability to classify traffic by
type or identify important events of interest from measured network traffic.
Over the past 15 years, machine learning models have become increasingly
integral to these tasks~\cite{nguyen2008survey, singh2013survey,
boutaba2018comprehensive}. Training a machine learning model from network
traffic typically involves extracting a set of features
that achieve good model performance, a process that requires
domain knowledge to know the features that are most relevant to prediction, as
well as how to transform those features in ways that result in separation
of classes in the underlying dataset. Figure~\ref{fig:traditional_cycle} shows a
typical pipeline, from measurement to modeling: The process begins with data
(\eg, a raw traffic trace, summary statistics produced by a measurement system);
features are then derived from this underlying data. The collection of features
and derived statistics is often referred to as the data {\em
representation} that is used as input to the model. Even for cases where the
model itself learns the best representation based on its input (\eg,
representation learning or deep learning), the designer of the algorithm must
still determine the {\em initial} representation of the data that is provided to
model. 

Unfortunately, with existing network traffic measurement systems, the first
three steps of this process---collection, cleaning, and feature
engineering---are often out of the pipeline designer's control. To date, most
network management tasks that rely on machine learning from network traffic
have assumed the data to be fixed or given, typically because decisions about
measuring, sampling, aggregating, and storing network traffic data are made
based on the capabilities (and constraints) of current standards and hardware
capabilities (\eg, IPFIX/NetFlow).  As a result, a model might be trained
with a sampled packet trace or aggregate statistics about network
traffic---not necessarily because that data representation would result in an
efficient model with good overall performance, but rather because the decision
about data collection was made well before any modeling or prediction problems
were considered.

Existing network traffic measurement capabilities capture either flow-level
statistics or perform fixed transformations on packet captures. First,
flow-based monitoring collects coarse-grained statistics (\eg, IPFIX/NetFlow or
collection infrastructure such as Kentik~\cite{kentik} and
Deepfield~\cite{deepfield}). These statistics are also often based on samples of
the underlying traffic~\cite{Estan:2002:NDT:633025.633056}. Conversely,
packet-level monitoring aims to capture traffic for specialized monitoring
applications~\cite{corelight} or triggered on-demand to capture some subset of
traffic for further analysis~\cite{zhu2015packet}. Programmable network hardware
offers potential opportunities to explore how different data representations can
improve model performance; yet, previous work on programmable hardware and
network data structures has typically focused on efficient ways to aggregate
statistics~\cite{liu2016one} (\eg, heavy hitter detection), rather than
supporting different data representations for machine learning models. In all of
these cases, decisions about data representation are made at the time of
configuration or deployment, {\em well before the analysis takes place}. Once
network traffic data is collected and aggregated, it is difficult, if not
impossible, to retroactively explore a broader range of data representations
that could potentially improve model performance.

A central premise of the work in this paper is introducing additional
flexibility into the first three steps of this pipeline for network management
tasks that rely on traffic measurements.  On the surface, raw packet traces would seem to be an appealing starting
point: Any network operator or researcher knows full well that raw packet
traces offer maximum flexibility to explore transformations and representations
that result in the best model performance.  Yet, unfortunately, capturing raw
packet traces often proves to be impractical on large networks as raw packet
traces produce massive amounts of data, introducing storage and bandwidth
requirements that are often prohibitive. Many controlled laboratory experiments
(and much past work) that demonstrate a model's accuracy turn
out to be non-viable in practice because the systems costs of deploying and
maintaining the model are prohibitive. An operator may ultimately need to
explore costs across state, processing, storage, and latency to understand
whether a given pipeline can work in its network.

Evaluation of a machine learning model for network management tasks must also
consider the operational costs of deploying that model in practice. Such an
evaluation requires exploring not only how data representation and models affect
model accuracy, but also the systems costs associated with different
representations. Sculley \etal{} refer to these considerations
as ``technical debt''~\cite{sculley2015hidden} and identified a number
of hidden costs that contribute to building the technical debt of ML-systems,
such as: unstable sources of data, underutilized data, use of generic packages,
among others. This problem is vast and complex, and this paper does not explore
all dimensions of this problem. For example, we do not investigate practical
considerations such as model training time, model drift, the energy cost of
training, model size, and many other practical considerations. In this regard,
this paper scratches the surface of systemization costs, which we believe
deserves more consideration before machine learning can be more widely deployed
in operational networks.

To lay the groundwork for more research that considers these costs, we develop and publicly release a
systematic approach to explore the relationship between different data
representations for network traffic and (1)~the resulting model performance as
well as (2)~their associated costs. We present \system{} (\cref{sec:system}), 
a
proof-of-concept reference system implementation designed to explore network
data representations and evaluate the systems-related costs of these
representations. To facilitate exploration, \system{} implements a processing
pipeline that performs passive traffic monitoring and in-network feature
transformations at traffic rates of up to 10~Gbps in software (\cref{sec:eval}).
The pipeline supports  capture and real-time transformation into a
variety of common feature representations for network traffic; we have designed
and exposed an API so that \system{} can be extended to define new
representations, as well. In addition to facilitating the transformations thesemselves,
\system{} performs profiling to quantify system costs, such as state and
compute, for each transformation, to allow researchers and operators to evaluate
not only the accuracy of a given model but the associated systems costs of the
resulting representation.

We use \system{} to demonstrate the value of jointly
exploring data representations for modeling and their associated costs for two
supervised learning problems in networking: video quality inference from
encrypted traffic and malware detection. We study two questions:
\begin{itemize}[leftmargin=*]
    \item {\em How does the cost of feature representation vary with network
        speeds?} We use \system{} to evaluate the cost of performing different
        transformations on traffic in real-time in deployed networks across
        three cost metrics: in-use memory (\ie, state), per packet processing
        (\ie, compute), and data volume generated (\ie, storage). We show that
        for the video quality inference models, state and storage requirements
        out-pace processing requirements as traffic rates increase
        (\cref{sec:profiling_video}).  Conversely, processing and
        storage costs dominate the systems costs for the malware detection
        (\cref{sec:profiling_malware}). 
        These results suggest that fine-grained
        cost analysis can lead to different choices for traffic representation
        depending on different model performance requirements and network
        environments.

    \item {\em Can systems costs be reduced without affecting model accuracy?}
        We show that different data transformations allow
        systems designers to make meaningful decisions involving systems costs
        and model performance. For example, we find that state 
        requirements can be significantly reduced for both problems without
        affecting model performance (\cref{sec:cost_accuracy_video} and
        \cref{sec:cost_accuracy_malware}), providing important opportunities for
        in-network reduction and aggregation.
\end{itemize}
\noindent
While it is well-known that {\em in general} different data representations
can both affect model accuracy and introduce variable systems costs, 
network research has left this area relatively under-explored.
Our investigation both constitutes an important re-assessment of previous
results 
and lays the groundwork for new directions in applying machine learning to network
traffic modeling and prediction problems. From a scientific perspective, our
work explores the robustness of previously published results. From a deployment
standpoint, our results also speak to systems-level deployment considerations,
and how those considerations might ultimately affect these models in practice,
something that has been often overlooked in previous work. Looking ahead, we
believe that incorporating these types of deployment costs as a primary model
evaluation metric should act as a rubric for evaluating models that rely on
machine learning for prediction and inference from network traffic.

\section{Joint Exploration of Cost and Model Performance}\label{sec:cost_performance}

Exploring the wide
range of possible data representations can help improve model performance within
the constraints of what is feasible with current network technologies. Doing so,
however, requires a system that enables {\em joint
exploration} of both systems cost and model performance.
To this end, this section highlight two important requirements 
needed to support exploration: (1)~the ability to flexibly define how features
are extracted from traffic; and (2)~integrated analysis of systems costs.

\subsection{Flexible Feature Extraction}

Different network inference tasks use different models, each of which may depend
on a unique set of features. 
Consider the task of inferring the quality of a video streaming application from
encrypted traffic (\eg, resolution). This problem is
well-studied~\cite{bronzino2019inferring,
mazhar2018real,mangla2019using,gutterman2019requet}. The task has been commonly
modeled using data representations extracted from different networking layers at
regular intervals (\eg, every ten seconds). For instance, Bronzino
\etal{}~\cite{bronzino2019inferring} grouped data representations from different
networking layers into different feature sets: Network, Transport, and
Application layer features. Network-layer features consist of lightweight
information available from observing network flows (identified by the IP/port
four-tuple) and are typically available in monitoring systems (\eg,
NetFlow)~\cite{kentik,deepfield}. Transport-layer features consist of
information extracted from the TCP header, such as end-to-end latency and packet
retransmissions. Such features are widely used across the networking space but
can require significant resources (\eg, memory) to collect from large network
links. Finally, application-layer metrics are those that include any feature
related to the application data that can be gathered by solely observing packet
patterns (\ie, without resorting to deep packet inspection). These features
capture a unique behavior of the application and have been designed specifically
for this problem.

\begin{figure}[t!]
    \begin{minipage}{\linewidth}
        \centering
        \begin{subfigure}[b]{0.48\linewidth}
            \centering
            \includegraphics[width=\linewidth]{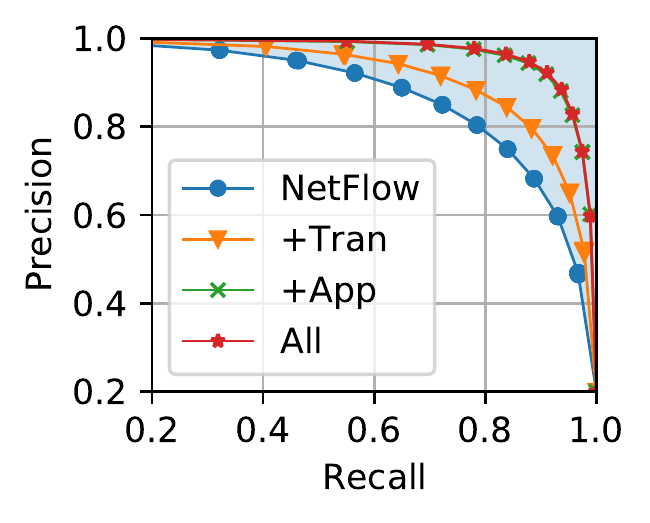}
            \caption{The relationship between data representations and model performance for video quality inference}
            \label{fig:performance}
        \end{subfigure} \hfill
        \begin{subfigure}[b]{0.48\linewidth}
            \centering
            \includegraphics[width=\linewidth]{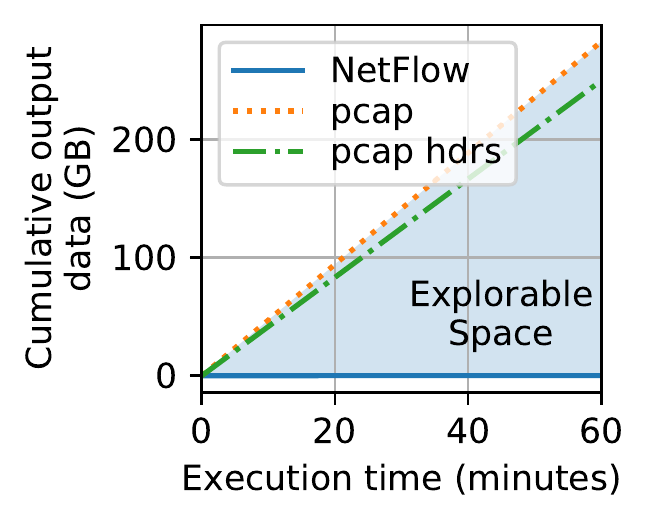}
            \caption{Storage cost for collecting a one hour of traffic across different monitoring system on a 10~Gbps link.}
            \label{fig:cost}
        \end{subfigure}
    \end{minipage}
    \vspace{-2mm}
    \caption{Balancing traffic data exploration and storage cost.}
    \label{fig:cost_performance}
  \end{figure}

We replicate the work of Bronzino \etal{}~\cite{bronzino2019inferring} by
training multiple machine learning models to infer the resolution of video
streaming applications over time using the three aforementioned data
representations. Figure~\ref{fig:performance} shows the precision and recall
achieved by each representation. We
observe that the performance of a model trained with Network Layer features
only (NetFlow in the figure) achieves the poorest performance, which agrees
with previous results. Hence, {\em relying
solely on features offered by existing network infrastructure would have produced
the worst performing models}. On the other hand, combining Network and
Application features results in more than a 10\% increase in both
precision and recall. This example showcases how limiting available data
representations to the ones typically available from existing systems (\eg,
NetFlow) can inhibit potential gains, highlighted
by the blue-shaded area in Figure~\ref{fig:performance}. This example
highlights the need for extensible data collection routines that can evolve
with Internet applications and the set of inference tasks.

\subsection{Integrated System Cost Analysis }

Of course, any representation is possible if packet
traces are the starting point, but raw packet capture can be prohibitive in
operational networks, especially at high speeds.  We demonstrate the amount of
storage required to collect traces at scale by collecting a one-hour packet
capture from a live 10~Gbps link. As shown in Figure~\ref{fig:cost}, we observe
that this generates almost 300~GB of raw data in an hour, multiple orders of
magnitude more than aggregate representations such as IPFIX/NetFlow. Limiting
the capture to solely storing packet headers reduces the amount of data
generated, though not enough to make the approach practical. To compute a
variety of statistics that would not be normally available from existing systems
we would require an online system capable of avoiding the storage 
requirements imposed by raw packet captures.

Deploying an online system creates practical challenges caused by
the volume and rate of traffic that must be analyzed. Failing to support
adequate processing rates (\ie, experiencing packet drops) ultimately degrades the
accuracy of the resulting features, potentially invalidating the models.
Fortunately, packet capture at high rates in software has become increasingly
feasible due to tools such as PF\_RING~\cite{deri2004improving} and
DPDK~\cite{dpdk}. Thus, in addition to exploiting the available technical
capabilities to {\em capture} traffic at high rates, the system should implement
techniques to maximize its ability to ingest traffic and lower the overhead
of system processing. For example, the system has to efficiently
limit heavyweight processing associated with certain features to subsets of
traffic that are targeted by the inference problem being studied without
resorting to sampling, which can negatively impact model performance.

Any feature transformation will introduce systems-related costs.  A network
monitoring system should make it possible to quantify the cost that such
transformations impose. Thus, to explore the space of model performance and their
associated systems costs, the system should provide an {\em integrated} mechanism to
profile each feature.

\begin{figure}[t!]
  \centering
  \includegraphics[width=\linewidth]{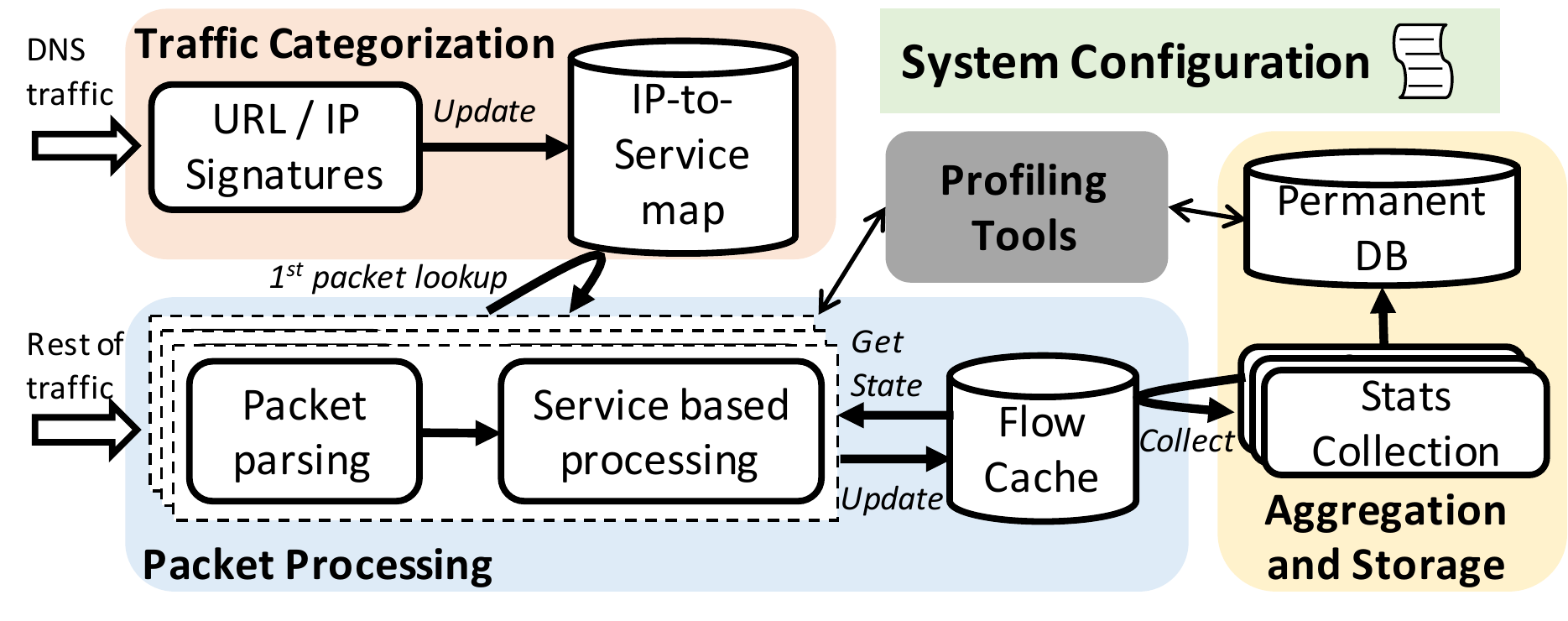}
  \vspace{-2mm}
  \caption{\sysname{} system overview.}
  \label{fig:systemoverview}
\end{figure}

\section{Exploring data representations with \sysname{}}\label{sec:system}

To explore network traffic feature representations and its subsequent effect
on both the performance of prediction models and collection cost, we need a
way to easily collect different representations from network traffic.  To
enable such exploration, we implement \sysname, which works both for
\textit{data representation design}, helping network operators explore the
accuracy-cost tradeoffs of different data representations for an inference
task; and for \textit{customized data collection in production}, whereby
\sysname\ can be deployed online to extract custom features.

Data representation design has three steps. First, network operators
or researchers define a superset of features worth exploring for the task and
configure \sysname{} to collect all these features for a limited time period.
Second, during this collection period, the system profiles the costs associated
with collecting each individual feature. Finally, the resulting data enables the
analysis of model accuracy versus traffic collection cost tradeoffs.

This section first describes the general packet processing pipeline of the
system (Section~\ref{sec:processing}) and how a user can configure this pipeline
to collect features specific to a given inference task
(Section~\ref{sec:extensibility}). We then
present how to profile the costs of features for data representation design
(Section~\ref{sec:profmethod}).

\subsection{Packet Processing Pipeline}\label{sec:processing}

Figure~\ref{fig:systemoverview} shows an overview of \sysname{}.
\sysname{} is implemented in Go~\cite{golang} to exploit 
performance and flexibility, as well as its built-in benchmarking tools. The
system design revolves around three guidelines: (1) Detect flows and
applications of interest early in the processing pipeline to avoid unnecessary
overhead; (2) Support state-of-the-art packet processing while minimizing the
entry cost for extending which features to collect; (3) Aggregate flow
statistics at regular time intervals and store for future consumption. The
pipeline has three components:
\begin{enumerate}
  \itemsep=-1pt
  \item a {\em traffic categorization} module responsible for associating
        network traffic with applications;
  \item a {\em packet capture and processing} module that collects network flow
        statistics and tracks their state at line rate; moreover, this block
        implements a cache used to store flow state information; and
  \item an {\em aggregation and storage} module that queries the flow cache to
        obtain features and statistics about each traffic flow and stores
        higher-level features concerning the applications of interest for later
        processing.
\end{enumerate}
\noindent
\system{} is customizable through a configuration file written in JSON. The
configuration provides a way to tune system parameters (\eg, which interfaces to
use for capture) as well as the definitions of service classes to monitor. A
service class includes three pieces of information that establish a logical
pipeline to collect the specified feature sets for each targeted service class:
(1)~which flows to monitor; (2)~how to represent the underlying flows in terms
of features; (3)~at what time granularity features should be represented.
Listing~\ref{lst:config} shows the JSON format used.

\begin{figure}[t!]
  \begin{lstlisting}[language=json, caption=Configuration example., captionpos=b, label={lst:config}]
    {
      "Name": "ServiceName",
      "(*@\hlred{Filter}@*)": {
        "DomainsString": ["domain.x", ...],
        "Prefixes": ["10.0.0.0/18", ...]
      },
      "(*@\hlblue{Collect}@*)": [FeatureSetA, FeatureSetB, ...],
      "(*@\hlyellow{Emit}@*)": 10
    }
  \end{lstlisting}
\end{figure}

\subsubsection{Traffic Categorization}

\sysname{} aims to minimize overhead generated by the processing and
state of packets and flows that are irrelevant for computing the features of
interest. Accordingly, it is crucial to categorize network flows based on their
service early so that the packet processing pipeline can extract features solely
from relevant flows, ideally without resorting to sampling traffic. To
accurately identify the sub-portions of traffic that require treatment online
without breaking encryption or exporting private information to a remote server,
\sysname{} implements a cache to map remote IP addresses to services accessed by
users. The map supports identifying the services flows belong to by using one of
two methods: (1)~{\em Using the domain name of the service:} similarly to the approach
presented by Plonka and Barford~\cite{plonka2011flexible}, \sysname{} captures
DNS queries and responses and inspects the hostname in DNS queries and matches
these lookups against a corpus of regular expressions for domain names that we
have derived for those corresponding services. For example,
$\verb/(.+?\.)?nflxvideo\.net/$ captures domain names corresponding to
Netflix's video caches. (2)~{\em Using exact IP prefixes:} For further flexibility,
\sysname{}~supports specifying matches between services and IP prefixes, which
assists with mapping when DNS lookups are cached or encrypted.

Using DNS to map traffic to applications and services may prove challenging in the
future, as DNS becomes increasingly transmitted over encrypted transport (\eg,
DNS-over-HTTPS~\cite{borgolte2019dns} or DNS-over TLS~\cite{reddy2017dns}). In
such situations, we envision \sysname{} relying on two possible solutions: (1)
parse TLS handshakes for the server name indication (SNI) field
in client hello messages, as this information is available in plaintext; or (2)
implement a web crawler to automatically generate an
IP-to-service mapping, a technique already implemented in production
systems~\cite{deepfield}.

\subsubsection{Packet Capture and Processing}

The traffic categorization and packet processing modules both require access to
network traffic. To support fast (and increasing) network speeds, \sysname{}
relies on state-of-the-art packet capture libraries: We implement \sysname{}'s
first two modules and integrate a packet capture interface based on
PF\_RING~\cite{deri2004improving} and the gopacket \emph{DecodingLayerParser}
library~\cite{gopacket}. \sysname{} also supports \emph{libpcap}-based packet
capture and replay of recorded traces.

Processing network traffic in software is more achievable than it has been in
the past; yet, supporting passive network performance measurement
involves developing new efficient algorithms and processes for traffic
collection and analysis. \sysname{} implements parallel traffic processing
through a pool of worker processes, allowing the system to scale capacity and
take advantage of multicore CPU architectures. We exploit flow
clustering (in software or hardware depending on the available resources) to
guarantee that packets belonging to the same flow are delivered to the same
worker process, thus minimizing cross-core communication and ensuring thread safety. The workers store the
computed state in a shared, partitioned flow cache, making it available for
quick updates upon receiving new packets.


The packet processing module has two components:

\paragraph{State storage: Flow cache.}
We implement a flow cache used to store a general data structure containing
state and statistics related to a network flow. The general flow data structure
allows storing different flow types, and differing underlying statistics using a
single interface. Furthermore, it includes, if applicable, an identifier to
match the services the flow belongs to. This information permits the system to
determine early in the pipeline whether a given packet requires additional
processing. The current version of the system implements the cache through a
horizontally partitioned hash map. The cache purges entries for flows that have
been idle for a configurable amount of time. In our configuration this timeout
is set to 10 minutes.

\paragraph{Feature extraction: Service-driven packet processing.}
A worker pool processes all non-DNS packets. Each worker has a dedicated capture
interface to read incoming packets. As a first step, each worker pre-parses MAC,
network, and transport headers, which yields useful information such as the
direction of the traffic flow, the protocols, and the addresses and ports of the
traffic. The system then performs additional operations on the packet depending
on the service category assigned to the packet by inspecting the flow's service
identifier in the cache. Using the information specified by the configuration
file, \sysname{} creates a list of feature classes to be collected for a given
flow at runtime. Upon receiving a new packet and mapping it to its service,
\sysname{} loops through the list and updates the required statistics.

\subsubsection{Aggregation and Storage}
\sysname{} exports high-level flow features and data representations at regular
time intervals. Using the time representation information provided in the
configuration file, \sysname{} initializes a timer-driven process that extracts
the information of each service class at the given time intervals. Upon firing
the collection event, the system loops through the flows belonging to a given
service class and performs the required transformation (\eg, aggregation or
sampling) to produce the data representation of the class.  \sysname{}'s packet
processing module exposes an API that provides access to the information stored
in the cache. Queries can be constructed based on either an application (\eg,
Netflix), or on a given device IP address. In the current version of the system,
we implement the module to periodically query the API to dump all collected
statistics for all traffic data representations to a temporary file in the
system. We then use a separate system to periodically upload the
collected information to a remote location, where it can be used as input to
models.

\subsection{User-Defined Traffic Representations} \label{sec:extensibility}

The early steps of any machine learning pipeline involve designing features
that could result in good model performance. We design \sysname{} to
facilitate the exploration of how different representations 
affect model performance and collection cost.
To 
do so, we design \sysname{} to use convenient flow abstraction
interfaces to allow for quick implementation of collection methods for features
and their aggregated statistics. Each flow data structure implements two
functions that  define how to handle a packet in the latter two steps of the
processing pipeline: (1)~an \texttt{AddPacket} function that defines how to
update the flow state metrics using the pre-processed information parsed from
the packet headers; and (2)~a \texttt{CollectFeatures} function that allows the user
to specify how to aggregate the features collected for output when the
collection time interval expires.

Implemented features are added as separate files.
\system{} uses the configuration file to obtain the list of service class
definitions and the features to collect for each one of them. Upon execution,
the system uses Go's language run-time reflection to load all available feature
classes and select the required ones based on the system configuration. The
implemented functions are then executed respectively during the packet
processing step or during representation aggregation.  We detail in
Section~\ref{sec:deployment} how the system can be configured to flexibly
collect features at deployment time for two use cases: video quality inference and malware detection. 

\begin{figure}[!t]
  \begin{lstlisting}[language=Go, caption=Implementing a new counters class., captionpos=b, label={lst:counters}]
    // PacketCounters is a data structure to collect packet and byte counters
    type PacketCounters struct {
      InCounter  int64
      OutCounter int64
      InBytes    int64
      OutBytes   int64
    }

    // AddPacket increment the counters based on information contained in pkt
    func (c *PacketCounters) AddPacket(pkt *network.Packet) {
      if pkt.Dir == network.TrafficIn {
        c.InCounter++
        c.InBytes += pkt.Length
      } else if pkt.Dir == network.TrafficOut {
        c.OutCounter++
        c.OutBytes += pkt.Length
      }
    }

    // PacketCountersOutput is a data structure that contains the features to output
    type PacketCountersOutput struct {
      KbpsUp    float64
      KbpsDw    float64
      PpsUp     float64
      PpsDw     float64
    }

    // CollectFeatures calculates the features to output at given time intervals
    func (c *PacketCounters) CollectFeatures(slotSize float64) PacketCountersOutput{
      return PacketCountersOutput{
        KbpsUp:  float64(c.OutBytes) / (slotSize * 128.0)
        KbpsDw:  float64(c.InBytes) / (slotSize * 128.0)
        PpsUp: float64(c.OutBytes) / slotSize
        PpsDw: float64(c.InBytes) / slotSize
      }
    }
  \end{lstlisting}
  \vspace{-2mm}
\end{figure}

As an example, we show in Listing~\ref{lst:counters} our implementation of the
PacketCounters feature class. This collection of features, stored in the
\texttt{PacketCounters} data structure, keeps track of the number of packets and
bytes for observed flows. To do so, the \texttt{AddPacket} function uses the
pre-processed information which is stored in the \texttt{Packet} provided as
input (\ie, the direction of the packet and its length). Upon triggering of the
collection interval, the system uses the structure to output throughput and
packets per-second statistics, \ie, the \texttt{CollectFeatures} function and
the output data structure \texttt{PacketCountersOutput}. The current release of
the system provides a number of built-in default features commonly collected
across multiple layers of the network stack. Table~\ref{tab:features} provides
an overview of the features currently supported.

\begin{table}[t!] \centering
  \resizebox{0.95\columnwidth}{!}{
    \begin{tabular}{ll}
      \toprule
     \textbf{Group} & \textbf{Features}                                \\ \midrule
      PacketCounters        & throughput, packet counts                        \\
      \rowcolor{Gray} PacketTimes           & packet interarrivals                             \\
      TCPCounters           & flag counters, window size, retransmissions, etc.\\
      \rowcolor{Gray} LatencyCounters       & latency, jitter                                  \\
      \bottomrule
    \end{tabular}}

  \caption{Current common features available in \sysname{}.}
  \vspace{-2mm}
  \label{tab:features}
\end{table}

\paragraph{Design considerations.}
We took this design approach to offer full flexibility in defining new features
to collect while also minimizing the amount of knowledge required of a user
about the inner mechanics of the system. We made several
compromises in developing \system.  First, our design focuses on supporting
per-flow statistics and output them at regular time intervals. This approach
enables the system to exploit established packet processing functions (\eg,
clustering) to improve packet processing performance. Conversely, this solution
might limit a user's ability to implement specific types of features, such as
features that require cross-flow information or those based on events. Second, the
software approach for feature calculation proposed in \system{} might
encourage a user
to compute statistics that are ultimately unsustainable for an online system
deployed in an operational network. To account for this possibility, the next section
discusses how the system's cost profiling method provides a way to quantify the
cost impact that each feature imposes on the system. Ultimately, this analysis
should provide feedback to a user in understanding whether such features should
be considered for deployment.

\subsection{Cost Profiling}\label{sec:profmethod}

\sysname{} aims to provide an
intuitive platform to evaluate the system cost effects of the user
defined data representations presented in the previous section. To do so, we
use Go's built-in benchmarking features and implement dedicated tools to
profile different costs intrinsic to the collection process. At data
representation design time, users employ the  profiling method to quickly
iterate through the collection of different features in isolation and provide a
fair comparison for three cost metrics: state, processing, and storage.

\paragraph{State costs.}
We aim to collect the amount of in-use memory over time for each feature class
independently. To achieve this, we use Go's \texttt{pprof} profiling tool. Using
this tool, the system can output at a desired instant a snapshot of the entire
in-use memory of the system. We extract from this snapshot the amount of memory
that has been allocated by each service class at the end of each iteration of
the collection cycle, \ie, the time the aggregation and storage module gathers the
data from the cache, which corresponds to peak memory usage for each interval.

\paragraph{Processing costs.}
To evaluate the CPU usage for each feature class, we aim to monitor the amount
of time required to extract the feature information from each packet, leaving
out any operation that shares costs across all possible classes, such as
processing the packet headers or reading/writing into the cache. To do so, we
build a dedicated time execution monitoring function that tracks the execution
of each \texttt{AddPacket} function call in isolation, collecting running
statistics (\ie, mean, median, minimum, and maximum) over time. This method is
similar in spirit to Go's built-in benchmarking feature but allows for using raw
packets captured from the network for evaluation over longer periods of time.

\paragraph{Storage costs.}
Storage costs can be compared by observing the size of the output generated over
time during the collection process. The current version of the system stores
this file in JSON format without implementing any optimization on the
representation of the extracted information. While this solution can provide a
general overview of the amount of data produced by the system, we expect that this
feature will be further optimized for space in the future. 

\paragraph{Cost profiling analysis.} 
\sysname\ supports two modes for profiling feature costs: (1) Profiling from
live traffic: in this setting the system captures traffic from a network
interface and collects statistics for a configurable time interval;  and (2)
Profiling using offline traffic traces: in this setting profiling runs over
recorded traffic traces, which enables fine-grained inspection of specific
traffic events (\eg, a single video streaming session) as well as repeatability
and reproducibility of results. Similarly to Go's built-in benchmarking tools,
our profiling tools run as standalone executables. To select the sets of
user-defined features (as described in Section~\ref{sec:extensibility}) to
profile, the profiling tool  takes as input the same system configuration file
used for executing the system. Upon execution, the system creates a dedicated
measurement pipeline that collects statistics over time. 


\section{\mbox{Prototype Evaluation}}\label{sec:eval}

To examine the traffic processing capacity of \sysname{}, we deploy the system
on a commodity server equipped with 16 Intel Xeon CPUs running at 2.4~GHz, and
64 GB of memory running Ubuntu 18.04. The server has a 10~GbE link that receives
mirrored traffic from an interconnect link carrying traffic for a consortium of
universities.\footnote{All captured traffic has been anonymized and sanitized to
obfuscate personal information before being used. No sensitive information has
been stored at any point. Our research has been approved by the university’s
ethics review body.} The link routinely reaches nearly full capacity (\eg,
roughly 9.8~Gbps) during peak times each day during the academic year. We evaluate
\sysname{} on the link over several days in October 2020. We use the
PF\_RING packet library with zero-copy enabled in order to access packets with
minimal overhead. We did not specifically engineer or optimize our
prototype for high-speed processing in production environments---such
optimization is not the focus of this paper---yet this setup nonetheless
offers a realistic deployment representation in an ISP network where an operator
might use \sysname{} to deploy its monitoring models, enables us to better
understand potential system bottlenecks, and demonstrates that a production deployment of
\system{} is feasible.

\begin{figure}[t!]
    \begin{minipage}{\linewidth}
        \centering
        \begin{subfigure}[b]{0.48\linewidth}
            \centering
            \includegraphics[width=\linewidth]{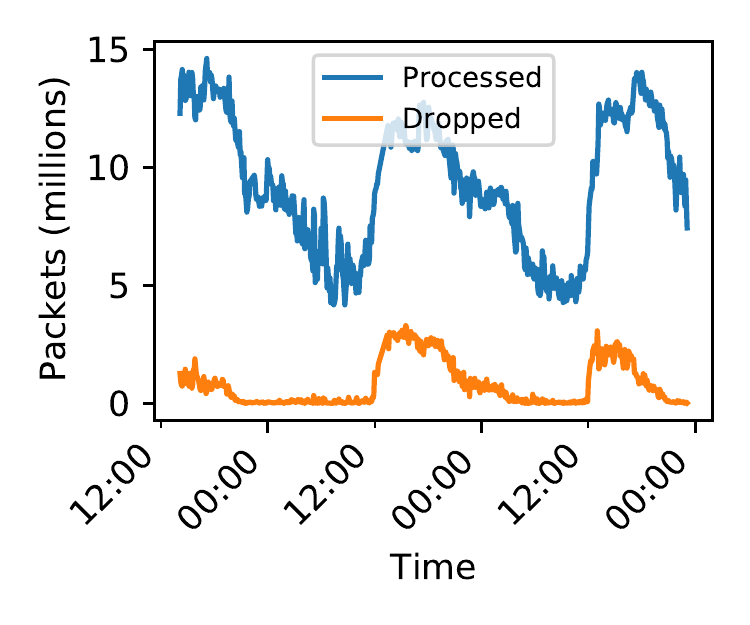}
            \caption{Packets processed vs drops}
            \label{fig:syscapacity}
        \end{subfigure} \hfill
        \begin{subfigure}[b]{0.48\linewidth}
            \centering
            \includegraphics[width=\linewidth]{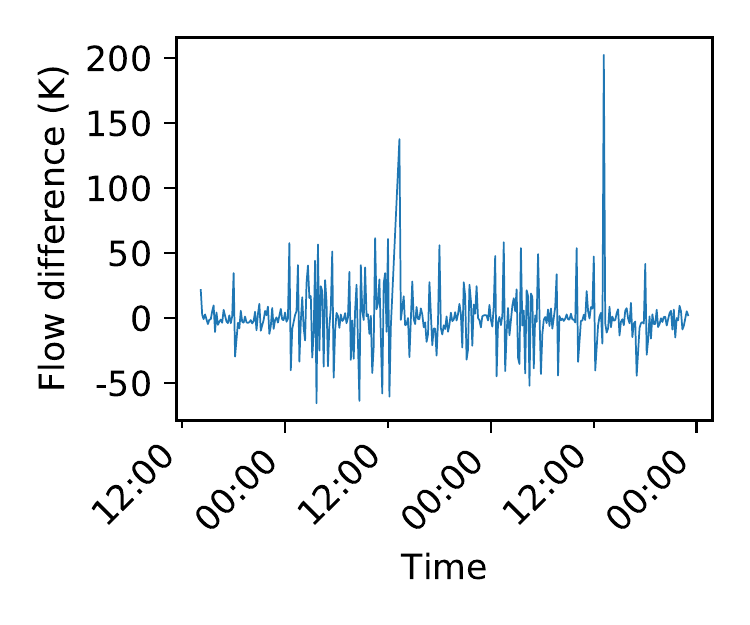}
            \caption{Flow cache size changes over time.}
            \label{fig:flowcachediff}
        \end{subfigure}
        \vspace{-2mm}
    \end{minipage}
    \caption{\sysname{} performance on the server.}
    \label{fig:system_performance}
\end{figure}

Figure~\ref{fig:syscapacity} shows the number of packets processed and the
number of packets dropped in 10-second windows over the course of a few days
collecting the features required to infer video quality metrics in real time for
eleven video services (more details on the use case are presented in the next
section). Traffic tends to show a sharp increase mid-day, which coincides with
an increase in the rate of packet drops. Overall, \sysname{} can process roughly
one million packets per-second (10M PPS per ten-second window in the figure)
without loss. Average packet size plays a significant role in the system's capacity;
for context, dShark~\cite{yu2019dshark} processes 3.3M PPS to process 40 Gbps by
assuming average packet size of 1500 bytes. More specialized hardware, or faster
general-purpose hardware could of course allow for for more parallel traffic
processing workers and better performance.

We investigate the cause of packet drops to understand
bottlenecks in \sysname{}. The system's flow cache is a central
component that is continuously updated concurrently by the workers that
process traffic. We study the ability of the system's flow cache to update the
collected entries upon the receipt of new incoming packets. We implement
benchmark tests that evaluate how many update operations the flow cache
can perform each second in isolation from the rest of the system. We test two
different scenarios: first, we evaluate the time to create a new entry in the
cache, \ie, the operation performed upon the arrival of a newly seen flow.
Second, we repeat the same process but for updates to existing flows in the
cache. Our results show that new inserts take one order of magnitude more time
than a simple update: roughly 6,000 nanoseconds (6 microseconds) versus 200 nanoseconds. 
Thus, the current flow cache implementation cannot support the creation of more than
about 150,000 new flows per second.

We confirm this result by looking at the arrival of new flows in our deployment.
Figure~\ref{fig:flowcachediff} shows the difference in the size of the flow
cache between subsequent windows over the observation period. Negative values
mean that the size of the flow cache decreased from one timestamp to the next.
As shown, there are sudden spikes (\eg, greater than 100,000 new flows) in the
number of flow entries in the cache around noon on two of the days, times that
correspond with increases in packet drops. Recall that the flow cache maintains
a data structure for every flow (identified by the IP/port four-tuple).
The spikes are thus a result of \sysname{} processing a large number of
previously unseen flows. This behavior helps explain the underlying causes for
drops. Packets for flows that are not already in the flow cache cause multiple
actions: First, \sysname{} searches the cache to check whether the flow already
exists. Second, once it finds no entries, a new flow object is created and
placed into the cache, which requires locks to insert an entry into the cache
data structure. We believe that performance might be improved
(\ie, drop rates could be lowered) by using a lock-free cache data structure and
optimizing for sudden spikes in the number of new flows. Such optimizations
are not the focus of this study, but we hope that our work lays the
groundwork for follow-up work in this area.


\section{\mbox{Use Cases}} \label{sec:deployment}

In this section, we use \system{} to
prototype two common inference tasks: streaming video
quality inference and malware detection.  For each problem, we conduct
 the three phases of the data representation design: (1)~definition and 
 implementation of a superset of candidate features; (2)~feature
collection and evaluation of system costs; and finally, (3)~analysis of the
cost-performance tradeoffs.

This exercise not only allows us to empirically measure systems-related costs of
data representation for these problems, but also to demonstrate that the
flexibility we advocate for developing network models is, in fact, achievable
and required in practice. Our analysis shows that in both use cases we can
significantly lower systems costs while preserving model performance. Yet, each
use case presents different cost-performance tradeoffs: the dominant costs for
video quality inference are state and storage, whereas for malware detection
they are online processing and storage. Further, our analysis demonstrates that
the ability to transform the data in different ways empowers systems designers
to take meaningful decisions at deployment time that affect both systems costs
as well as model performance.


\subsection{Video Quality Inference Analysis}\label{sec:video}

Video streaming quality inference often rely on features
engineered by domain experts~\cite{bronzino2019inferring,
mazhar2018real,mangla2019using,gutterman2019requet}. We
evaluate the models proposed by Bronzino
\etal~\cite{bronzino2019inferring}. 

\subsubsection{\system{} Customization}
As discussed in Section~\ref{sec:cost_performance}, Bronzino
\etal~\cite{bronzino2019inferring} categorized useful features for video quality
inference into three groups that correspond to layers of the network stack:
Network, Transport, and Application Layer features. In their approach, features
are collected at periodic time intervals of ten seconds. The first ten seconds
are used to infer the startup time of the video, while remaining time intervals
are used to infer the ongoing resolution of the video being streamed. 

We add approximately 100 lines of Go code to implement in \system{} the feature
calculation functions to extract application features (\ie,
\texttt{VideoSegments}). Further, we use built-in feature classes to collect
network (\ie, \texttt{PacketCounters}) and transport (\ie, \texttt{TCPCounters})
features. We use these classes to configure the feature collection for 11 video
services, including the four services studied in~\cite{bronzino2019inferring}:
Netflix, YouTube, Amazon Prime Video, and Twitch. We report a complete
configuration used to collect Netflix traffic features in
Appendix~\ref{app:video}. 
This use case, demonstrates how \system{} can be
easily used to collect common features (\eg, flow counters collected in NetFlow)
as well as extended to collect specific features useful for a given inference
task.

\subsubsection{\mbox{Data Representation Costs}}\label{sec:profiling_video}

\begin{figure*}[t!]
  \begin{minipage}{\linewidth}
  \begin{subfigure}[b]{0.32\linewidth}
    \centering
    \includegraphics[width=.92\linewidth]{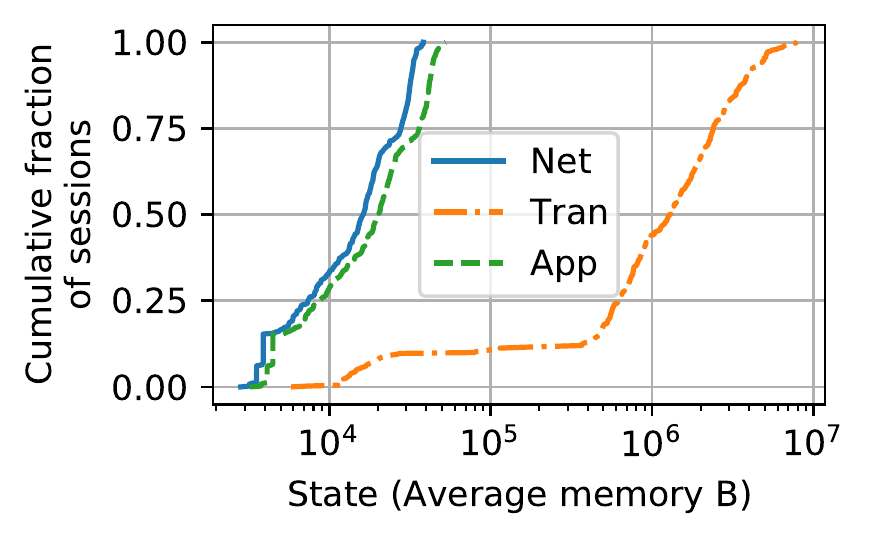}
    \vspace{-2mm}
    \caption{State required for different features.}
    \label{fig:mem_profile}
  \end{subfigure} \hfill
  \begin{subfigure}[b]{0.32\linewidth}
    \centering
    \includegraphics[width=.92\linewidth]{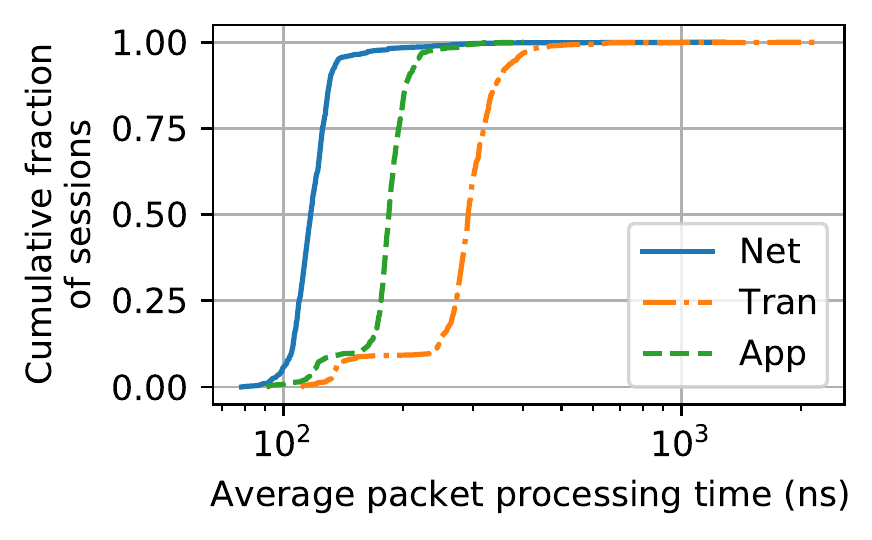}
    \vspace{-2mm}
    \caption{Processing costs for different representations.}
    \label{fig:cpu_profile}
  \end{subfigure} \hfill
  \begin{subfigure}[b]{0.32\linewidth}
    \centering
    \includegraphics[width=.92\linewidth]{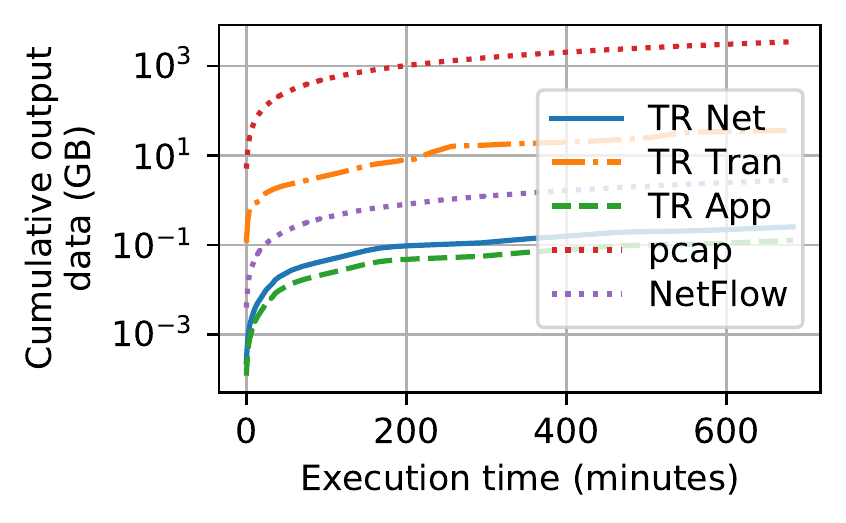}
    \vspace{-2mm}
    \caption{Storage costs for different representations.}
    \label{fig:storage_profile}
  \end{subfigure}
  \end{minipage}
  \vspace{-3mm}
  \caption{Cost profiling for video inference models.}
\end{figure*}

We evaluate system-related costs of the three classes of features used for video
quality inference problem: network, transport, and application features. First
we use \system's profiling tools to quantify the fine-grained costs imposed by
tracking video streaming sessions. To do so, we profile the per-feature state
and processing costs for pre-recorded packet traces with 1,000 video streaming
sessions split across four major video streaming services (Netflix, YouTube,
Amazon Prime Video, and Twitch). Then, we study the effect of collecting the
different classes of features at scale by deploying the system in a 10~Gbps
interconnect link.

We find that while some features add relatively little state (\ie, memory) and
long-term storage costs, others require substantially more resources.
Conversely, processing requirements are within the same order of
magnitude for all three classes of features.

\paragraph{State Costs.}
We study the state costs as the amount of in-use memory required by the system
at the end of each collection cycle---\ie, the periodic interval at which the
cache is dumped into the external storage. Figure~\ref{fig:mem_profile} shows
the cumulative distribution of memory in Bytes across all analyzed video
streaming sessions. The reported results highlight how collecting transport
layer features can heavily impact the amount of memory used by the system. In
particular, collecting transport features can require up to
three orders of magnitude more memory compared to network and application
features. Transport features require historical flow information (\eg, all
packets) in contrast with network features that require solely simple counters.

Further, the application features require a median of a few hundred
MB in memory on the monitored link, with a slightly larger memory footprint than
network features. At first glance, we assumed that this additional cost was due
to the need for keeping in memory the information about video segments being
streamed over the link. Upon inspection, however, we realized that streaming
protocols request few segments at a time per time slot (across the majority of
time slots the number of segments detected was lower than three), which leads to
a minimal impact on memory used. We then concluded that this discrepancy was
instead due to the basic Go data structure used to store video segments in
memory, \ie, a slice, which requires extra memory to implement its
functionality.

\paragraph{Processing Costs.}
Collecting features on a running system measuring real traffic provides the
ability to quantify the processing requirements for each target feature class.
We represent the processing cost as the average processing time required to
extract a feature set from a captured packet. Figure~\ref{fig:cpu_profile} shows
distributions of the time required to process different feature classes. 
Collecting simple network counters requires the least processing
time, followed by application and transport features.

While there are differences among the three classes, the
difference is relatively small and within the same order of magnitude. These
results highlight how all feature classes considered for video inference are
relatively lightweight in terms of processing requirements. Hence, for this
particular service class, state costs have a much larger impact than processing
cost on the ability of collecting features in an operational network.

\paragraph{Storage Costs.}
Feature retrieval at scale can generate high costs due to the need to move
the collected data out of the measurement system and to the location where it
will be ingested for processing. Figure~\ref{fig:storage_profile} shows the
amount of data produced by \sysname{} when collecting data for the three feature
classes relevant to the video streaming quality inference on the monitored link.
For comparison, we also include the same information for two different
approaches to feature collection: (a) pcap, which collects an entire raw packet
trace; (b) NetFlow configured using defaults (\eg, five minutes sampling), which
collects aggregated per flow data volume statistics; this roughly corresponds to
the same type of information collected by \sysname{} for the network layer
features (\ie, TR Net in the figure).

Storage costs follow similar trends as the state costs
previously shown. This is not surprising as the exported information is a
representation of the state contained in memory. More interesting outcomes can
be observed by comparing our system output to existing systems. Raw
packet traces generate a possibly untenable amount of data and if used
continuously can quickly generate terabytes of data. This result supports our
claim that collecting traces at scale and for long periods of time quickly
becomes impractical. Next, we notice that, even if not optimized, our current
implementation produces less data than NetFlow, even when exporting similar
information, \ie, network features. While this result mostly reflects the
different verbosity levels of the configurations used for each system, it
confirms that having additional flexibility in exporting additional features,
\eg, video segments information, may introduce low additional cost. In the next
section, we demonstrate that having such features available may result in
significant model performance benefits.

\subsubsection{Model Performance}\label{sec:cost_accuracy_video}

In this section, we study the relationship between model performance and system
costs for online video quality inference. We use previously developed models but
explicitly explore how {\em data representation} affects model performance. We
focus on state-related costs (\ie, memory), as for video quality inference,
state costs mirror storage costs and the differences in processing costs of the
feature classes is not significant (Section~\ref{sec:profiling_video}).
Interestingly, we find that the relationship between state cost and model performance
is not proportional. More importantly, we find that it is often possible to
significantly reduce the state-related requirements of a model without
significantly compromising prediction performance, further bolstering the case
for systems like \system{} that allow for flexible data representations.

\begin{figure}[t!]
  \begin{minipage}{\linewidth}
  \centering
  \begin{subfigure}[b]{0.49\linewidth}
    \centering
  \includegraphics[width=\linewidth]{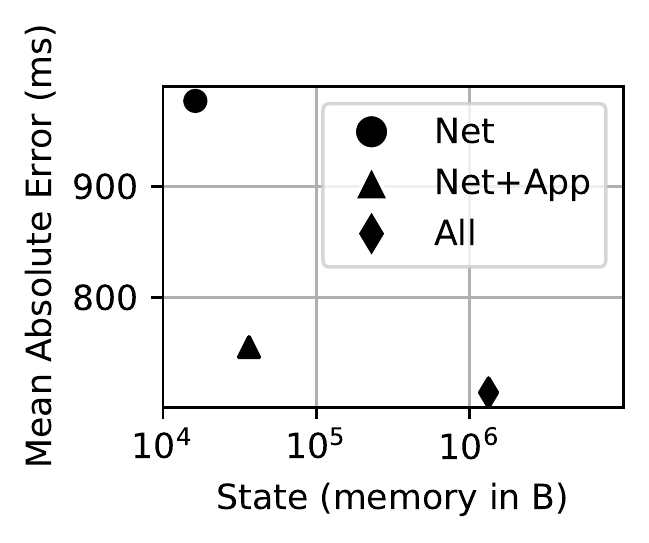}
  \vspace{-2mm}
  \caption{Startup delay (lower is better).}
  \label{fig:startup_features}
  \end{subfigure} \hfill
  \begin{subfigure}[b]{0.47\linewidth}
    \centering
    \includegraphics[width=\linewidth]{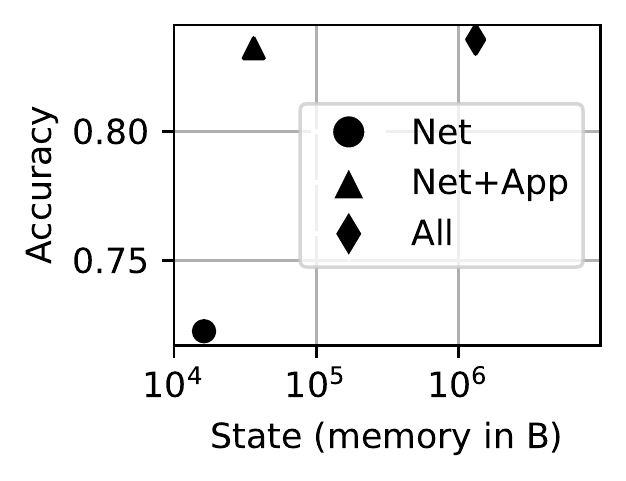}
    \vspace{-2mm}
    \caption{Resolution (higher is better).}
    \label{fig:resolution_features}
    \end{subfigure} \hfill
  \end{minipage}
  \vspace{-2mm}
  \caption{The relationship between features state cost and model performance for video streaming quality inference (marker shapes identify layers used and colors identify time interval size).}
  \label{fig:features_tradeoff}
\end{figure}

\paragraph{Representation vs. Model Performance.}
To understand the relationship between memory overhead and inference accuracy,
we use the dataset of more than 13k sessions presented
in~\cite{bronzino2019inferring} to train six inference models for the two
studied quality metrics: startup delay and resolution. For our analysis, we use
the random forest models presented in~\cite{bronzino2019inferring}; in
particular, random forest regression for startup delay and random forest
multi-class classifier for resolution. Further, we use the same inference
interval size, \ie, ten-second time bins.

Figure~\ref{fig:features_tradeoff} shows the relationship between model
performance and state costs. As shown, network features alone can provide
a lightweight solution to infer both startup delay and resolution but this
yields the lowest model performance. Adding application layer features
contributes to a very small additional memory overhead. This result is
particularly important for resolution where models with video segments alone
perform basically as well as combining all others. Further, adding
transport features (labeled ``All'' in the figure) provides limited benefits
in terms of added performance---40~ms on average lower errors for startup delay
and less than 0.5\% higher accuracy for resolution. Even for startup delay where
using transport features can improve the mean absolute error by a larger margin,
this comes at the cost of two orders of magnitude higher memory usage.

\begin{figure}[t!]
  \begin{minipage}{\linewidth}
  \centering
  \begin{subfigure}[b]{0.48\linewidth}
    \centering
    \includegraphics[width=\linewidth]{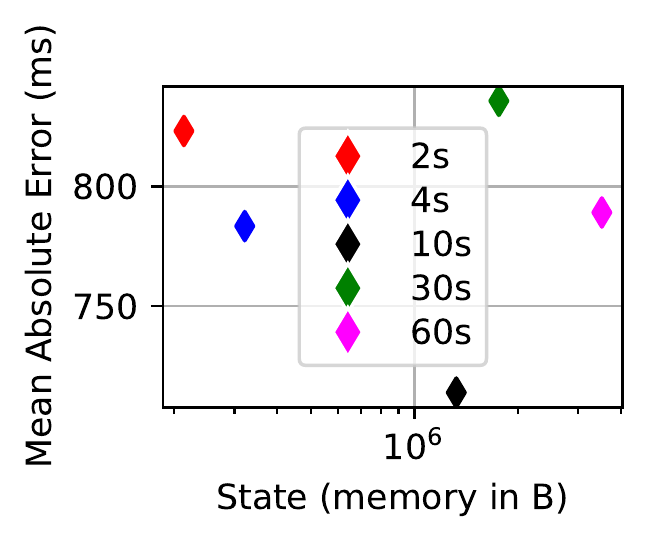}
    \vspace{-2mm}
    \caption{Startup delay (lower is better).}
    \label{fig:startup_bins}
    \end{subfigure}
  \begin{subfigure}[b]{0.48\linewidth}
    \centering
  \includegraphics[width=\linewidth]{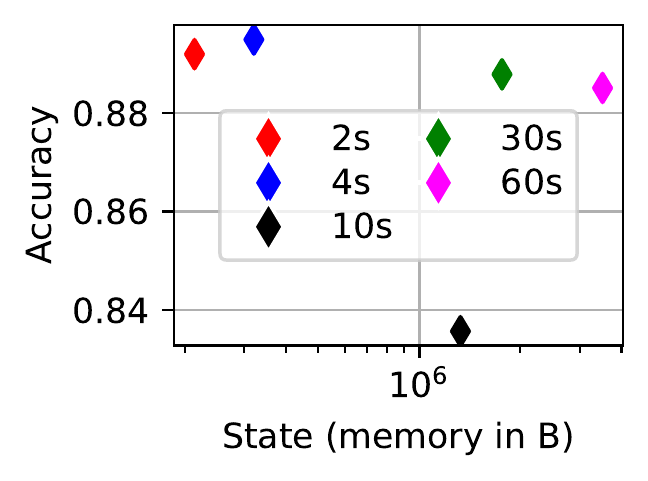}
  \vspace{-2mm}
  \caption{Resolution (higher is better).}
  \label{fig:resolution_bins}
  \end{subfigure}
  \end{minipage}
  \vspace{-2mm}
  \caption{The relationship between time granularity state costs and model performance for video quality inference (marker shapes identify layers used and colors identify time interval size).}

  \label{fig:time_tradeoff}
\end{figure}

\paragraph{Time Granularity vs. Model Performance.}
State of the art inference techniques (\ie, Bronzino
\etal~\cite{bronzino2019inferring} and Mazhar and Shafiq~\cite{infocomPaper})
employ ten-second time bins to perform the prediction of the features. This
decision is justified as a good tradeoff between the amount of information that
can be gathered during each time slot, \eg,~to guarantee that there is at least
one video segment download in each bin, and the granularity at which the
prediction is performed. For example, small time bins---\eg,~two seconds--- can
have a very small memory requirement but might incur in lower prediction
performance due to the lack of historical data on the ongoing session. On the
other hand, larger time bins---\eg,~60 seconds---could benefit from the added
information but would provide results that are just an average representation of
the ongoing session quality. These behaviors can be particularly problematic for
startup delay, a metric that would benefit from using exclusively the
information of the time window during which the player is retrieving data before
actually starting the video reproduction.

We train different random forest models
with increasing time bin sizes of 2, 4, 10, 30, and 60 seconds. To understand
the possible memory impact of the different time bins, we use all features (All)
to train the models. In Figure~\ref{fig:time_tradeoff}, we observe different
outcomes for the two quality metrics. For startup delay, the results show that
ten-second windows can indeed provide a good tradeoff between memory and
prediction accuracy,  achieving a minimum of 70~ms better predictions than all
other time granularities. This result shows that ten seconds is an acceptable
tradeoff between gathering enough information at the beginning of a session
without adding too much data from segment downloads that happens after the video
has started.

Interestingly, the results for resolution inference show that ten-second windows
perform the worst among all studied cases. This might be the product of multiple
factors. In particular, the change of the inference window size
not only changes how much information is used for prediction but it also
affects the granularly of the inference, possibly modifying the underlying
problem. Among the different time bin sizes we have different extremes ranging
from two-second windows, which are about the length of the shortest video
segments across all services, as well as 60-second time windows, which could
contain many video quality changes within the time slot caused by the download
of multiple video segments.

\begin{figure*}[t!]
  \begin{minipage}{\linewidth}
  \begin{subfigure}[b]{0.32\linewidth}
    \centering
    \includegraphics[width=.92\linewidth]{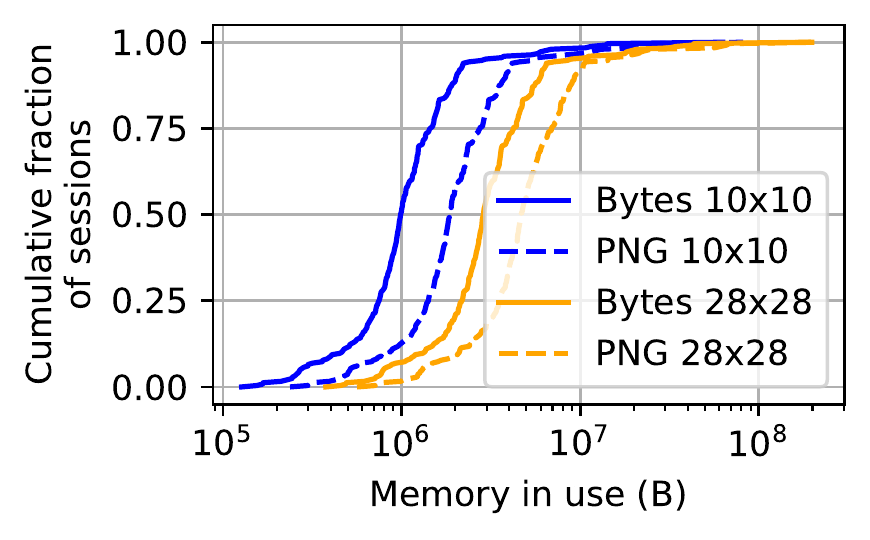}
    \vspace{-2mm}
    \caption{State required for different features.}
    \label{fig:mem_profile_malware}
  \end{subfigure} \hfill
  \begin{subfigure}[b]{0.32\linewidth}
    \centering
    \includegraphics[width=.92\linewidth]{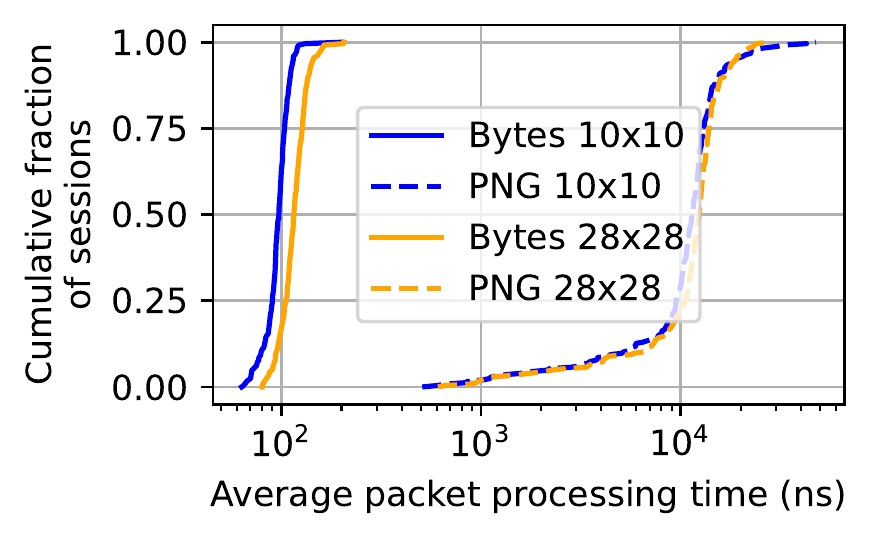}
    \vspace{-2mm}
    \caption{Processing costs for different representations.}
    \label{fig:cpu_profile_malware}
  \end{subfigure} \hfill
  \begin{subfigure}[b]{0.32\linewidth}
    \centering
    \includegraphics[width=.92\linewidth]{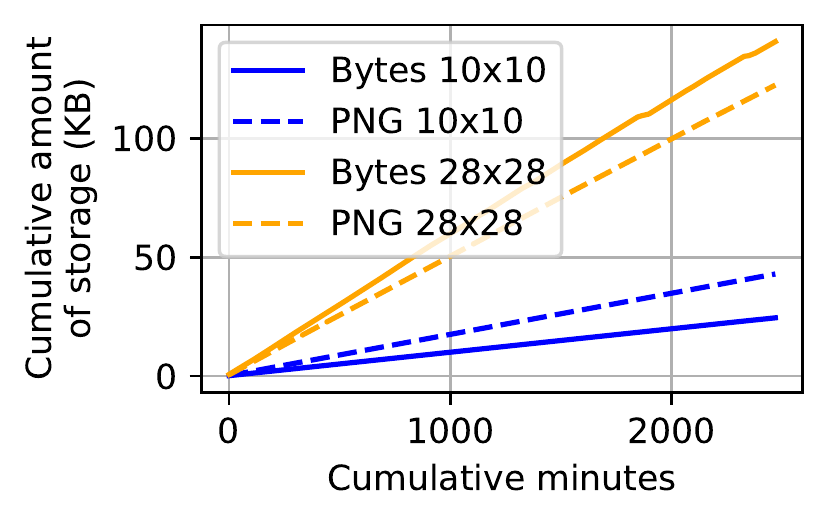}
    \vspace{-2mm}
    \caption{Storage costs for different representations.}
    \label{fig:storage_profile_malware}
  \end{subfigure}
  \end{minipage}
  \vspace{-3mm}
  \caption{Cost profiling for the malware detection models.}
  \label{fig:profile_malware}
\end{figure*}

\begin{figure}[t!]
  \begin{minipage}{\linewidth}
  \centering
  \begin{subfigure}[b]{0.49\linewidth}
    \centering
    \includegraphics[width=\linewidth]{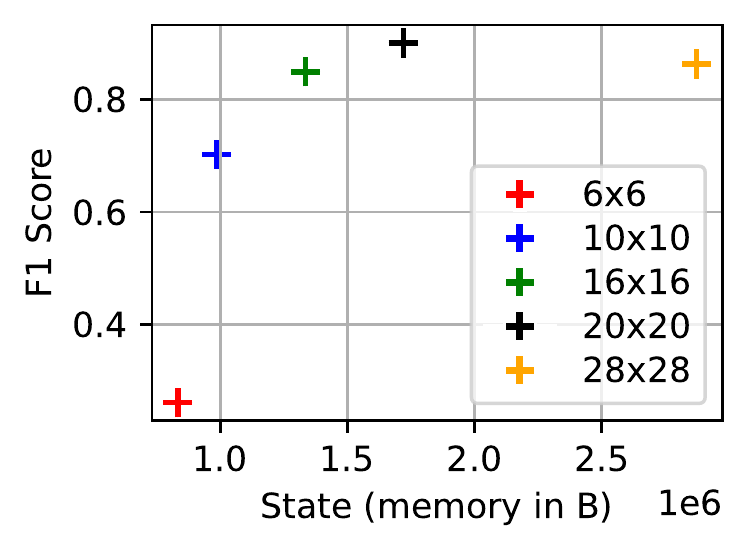}
    \vspace{-2mm}
    \caption{Image size (higher is better).}
    \label{fig:malware_mem_f1}
  \end{subfigure}
  \begin{subfigure}[b]{0.47\linewidth}
    \centering
    \includegraphics[width=\linewidth]{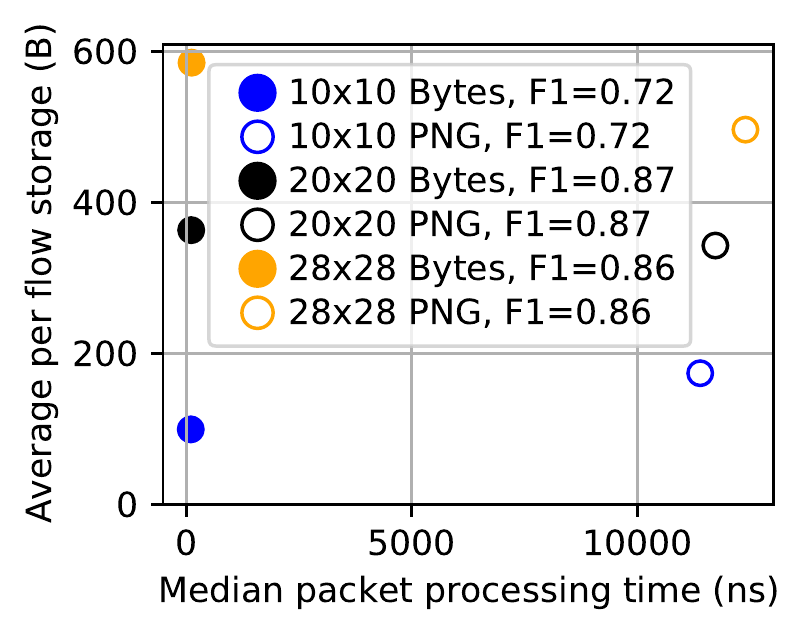}
    \vspace{-2mm}
    \caption{Impact of method on costs.}
    \label{fig:proc_stor_tradeoff}
  \end{subfigure}
  \end{minipage}
  \caption{The relationship between image size, processing method, and model performance for malware detection (marker shapes identify the method used and colors identify image size).}

  \label{fig:time_tradeoff}
\end{figure}

\subsection{Malware Detection Analysis}

In recent years, several works on traffic classification explored the
application of deep learning on raw network traffic to solve a variety of
inference tasks, such as malware
detection~\cite{marin2018rawpower,wang2017malware, marin2020deepmal,
wang2017hast} and service
identification~\cite{DBLP:journals/corr/abs-2008-02695}. Deep-learning based
solutions differ from alternative approaches (\eg, the one used for the
video inference in the previous section) in that they do not require to
determine the initial representation of the data that is provided to the model
but rather let the model learn the best representation based on its input. In
practice, this means that applying such classifiers online requires to feed a
Convolutionary Neural Network (CNN) with raw traffic data, represented either as
a normalized sequence of bytes~\cite{marin2018rawpower,marin2020deepmal} or by
converting the bytes into a gray-scale image~\cite{wang2017malware,
wang2017hast}.

The reduced complexity of these methods implies that the model and system
designers' role is limited to three choices: (1) the size of the input data
collected from the traffic (\eg, for traffic flow classification, Wang et
al.~\cite{wang2017malware} use the first 784 bytes of each flow, whereas
DeepMAL~\cite{marin2020deepmal} uses the first 100 bytes of payload of the first
two packets of each flow); (2) the layers to collect such data from (\eg, packet
headers and/or payload); and finally, (3) whether to perform data transformation
(\eg, produce a PNG image from raw bytes) online or offline. In this section, we
explore the impact of these different decisions on the deployment costs and
model performance. We find that depending on the cost metrics under study there
may not be a single ``best'' data representation, further bolstering the case
for systems like \system{} that allow for flexible data representations
exploration.

\subsubsection{\system{} Customization}
We add approximately 150 lines of Go code to implement both transformation methods in \system{}. For the first case, we implement a \texttt{BytesCopyCounters} data type that stores raw data extracted in a bytes array. Upon receiving new packets, the \texttt{AddPacket} function checks how much data is left to copy and, if any, it copies raw bytes from the correct layers into memory (\ie, headers, payload, or both). Once the bytes array is fully formed, no more packets are treated. Finally, the array is flushed into the output file when the collect function is invoked. For the second case, we implement a \texttt{PNGCopyCounters} data type that collects raw bytes from collected packets in a similar fashion but, also, converts the bytes into a PNG data structure when enough data has been collected. The data structure is flushed into the output file when the collect function is invoked.

\subsubsection{\mbox{Data Representation Costs}}~\label{sec:profiling_malware}
Similarly to the video inference use case, we use \system's profiling tools to
quantify the fine-grained costs imposed by the data collection required for deep
neural network malware detection. For this task, we use the CIC-IDS2017
dataset~\cite{sharafaldin2018toward}, a standard dataset used for malware
detection training and testing. The dataset consists of five days (working hours
only) worth of pcap traces that contain a mix of lab-generated benign and
malware traffic, for a total of \textasciitilde171k flows distributed across the
dataset. We divide the traces into ten minutes traffic sessions and profile the
state, processing, and storage costs across different configurations.

We focus our presentation on two of the three design factors that affect
possible configurations: (1) the data input size used to create the input image; 
and (2) whether to perform online data transformations. We do not present
results on the impact of using different parts of the packets (\ie,
headers, payload, or both) because our tests show that this
configuration has relatively small impact on the cost profiles. For all
presented results we solely use information collected from packet headers as
recent work demonstrated that this is often sufficient for most inference
tasks~\cite{DBLP:journals/corr/abs-2008-02695}.

\paragraph{State Costs.}
Figure~\ref{fig:mem_profile_malware} shows the
cumulative distribution of memory in Bytes across all analyzed traffic sessions
for different image sizes and techniques. As expected, the
results follow the base intuition that state directly relates to the size of the
memory allocated for each bytes array. Further, storing PNG images in memory
causes state costs to roughly double, due to the need for allocating memory for
both the raw bytes of memory as well as for the image.

\paragraph{Processing Costs.}
Figure~\ref{fig:cpu_profile_malware} shows distributions of the time required to
process the input data for different configurations. Storing raw
bytes in memory has a small impact on processing time, especially compared to
the online creation of PNG images (a two orders of magnitude larger median).
This behavior is expected since generating PNG images requires processing raw
bytes through multiple filtering and compression stages~\cite{png}.

Note that in this use case, most of the processing happens within the first few packets of each flow, which is the number of packets needed to achieve the desired input size for the CNN model. For this reason, flow size greatly impacts the average packet processing time where the majority of packets of a short flow are retained for processing, whereas the majority of packets of a large flow are not retained, and thus have negligible processing cost. To quantify this, we compute the average processing time for
one session of the dataset while dividing encountered flows based on their size,
\ie, in terms of the number packets that belong to the flow. For example, 
small flows (\ie, five packets or less) have an average processing
time of 13,060~ns for the 28x28 PNG transformation case. In contrast, the average processing time for large
flows (100 packets or more) is 335~ns. We observe similar results for all
configurations.

\paragraph{Storage Costs.}
Figure~\ref{fig:storage_profile_malware} shows the cumulative amount of data
produced by \sysname{} when collecting data for different configurations over
the entire dataset. Interestingly, while for small image sizes
(\eg, 10x10) PNG encoding generates higher volumes of data, for larger ones
(\eg, 28x28) it can save up to \textasciitilde20\% of total storage when
compared to saving in raw bytes. This reduction happens thanks to the compression
algorithms employed during the PNG encoding process. Compression is more effective
when images contain solely packet headers, as multiple fields repeat across packets. 
This observation
suggests that depending on the configuration employed in the system deployment,
different strategies might be adopted regarding where to transform raw bytes
into images.

\subsubsection{Model Performance}\label{sec:cost_accuracy_malware}

To understand the relationship between costs and inference accuracy, we use the
\textasciitilde171k flows contained in the CIC-IDS2017
dataset~\cite{sharafaldin2018toward} to train a CNN to classify traffic flows as
either benign or malware. We follow the approach of Wang et al.~\cite{wang2017malware}
by converting the sequence of bytes extracted into gray image PNGs before
feeding them to the CNN; hence, we refer to the size of the sequence of bytes as
the image size. We train 15 models in total using different
configurations for image size (6x6, 10x10, 16x16, 20x20, and 28x28) and the
layers of the network stack used to extract the sequence of bytes (header-fields
only, payload only, or header-fields and payload). We observe  that using
payload only yields the lowest F1 score, while the other two configurations yield similar results 
(within a 3\%  F1 score difference). Given these small differences, we focus 
the remainder of this  section on results varying configuration sizes alone. 
We perform the cost-performance analysis on a test set of 19k flows. 

\paragraph{Representation vs. Model performance.}
We explore the relationship between image size and inference performance.
Figure~\ref{fig:malware_mem_f1} presents the F1 score obtained as we vary the
image size. We present results solely for raw bytes, as F1 scores and cost trends 
are similar between the two methods. Image size has a clear impact on the model performance and the
associated memory cost. The F1 score improves as the image size increases.
However, this improvement flattens as the image size increases from a 20x20 to
28x28, whereas the induced cost in terms of memory cost maintained per-flow
doubles. This result supports the need to explore the performance-cost
relationship, as maintaining more information does not always lead to higher
performance.

Following the results obtained in the previous section, we explore whether
different configurations can lead to different tradeoffs and, consequently,
different deployment strategies. Figure~\ref{fig:proc_stor_tradeoff} shows the
storage costs in terms of average storage per flow generated against the average
processing time for different configurations. While processing
times are consistently orders of magnitude higher when the PNG generation is
integrated within the processing pipeline, storage costs have the opposite trend
for larger image configurations, which are also the best performing ones. This
result suggests that while converting an image can be computationally unfeasible
under certain deployment settings, it might be preferable to perform it online
for scenarios where storage is a major constraint.

%


\section{Related Work}

Machine learning models have become 
integral for solving many network management tasks~\cite{nguyen2008survey, singh2013survey, boutaba2018comprehensive}, 
from performance inference to security.
Collecting input data to build models for network management tasks can be
typically achieved with passive network monitoring tools, such as packet
captures (\eg, libpcap~\cite{pcap} and its derivative applications
Wireshark~\cite{orebaugh2006wireshark} and Tshark~\cite{www-tshark}) or flow
captures (\eg, NetFlow~\cite{claise2004cisco}, IPFIX~\cite{ipfix}).
Unfortunately, this set of network monitoring tools inhibits model designers
from exploring the space of possible data representations. On one hand, packet
captures generate massive volume of data which makes them a none-viable approach
for large networks. On the other, flow captures produce statistical information
that are too coarse grained to enable full exploration of all possible data
representations. Similarly, streaming analytics
platforms~\cite{cranor2003gigascope, 180224, Yuan:2017:QNM:3098822.3098830,
narayana2017language, Gupta:2018:SQS:3230543.3230555} and algorithms (\eg,
``sketches'')~\cite{yu2013software,kumar2004data, liu2016one, yang2018elastic}
allow operators to express queries on streaming traffic data but they are
primarily designed to collect low-level statistics on a backbone router or
switch, or a programmable datacenter switch, which operate at very high speeds.
As such, they typically support a more limited set of queries that are
constrained by the hardware they are designed to support.

To obviate to the limitations of packet capture tools at scale,
dShark~\cite{yu2019dshark} implements a distributed computing engine for
processing distributed network traces, at scale, in the data center. As in
\sysname{}, dShark offers a programming interface that permits (1)~declaring
groups of packet summaries that have similar properties and (2)~defining queries
that operate on such packets. \system{} differs in its design as it focuses on
transformations as input to machine learning models. Conversely, dShark focuses its
design on multi-point collection and transformations required for
general diagnosis. Further, \sysname{} focuses on helping
model designers evaluate both model accuracy and systems-related costs
associated with data collection.

Advanced network monitoring and analysis tools such as
Tstat~\cite{mellia2003tstat,finamore2010live}, Bro~\cite{paxson1999bro}, and
Snort~\cite{roesch1999snort} are closest in spirit to~\sysname{} in that they
have the goal of capturing network traffic and executing transformations on
the data for later use. Tstat is an open source passive monitoring tool that
can monitor network traffic and output logs, statistics, and histograms with
different granularities: per-packet, per-flow, or aggregated. Bro and Snort
are network intrusion detection systems that rely on regular expressions to
identify the subset of packets to inspect and execute specific tasks based on
the class of traffic. Ultimately, these tools would need to be adapted to
achieve custom feature representation, data representation exploration, and
profiling data collection costs. Some commercial products apply machine
learning to network traffic (\eg, Nokia's Traffica~\cite{traffica},
Deepfield~\cite{deepfield}, NIKSUN's NetCVR~\cite{netvcr}); these approaches
are are proprietary and address a specific problem (e.g., customer support).
On the other hand, \sysname{} is open-source, and permits jointly evaluating
model performance and features collection costs at design time.

Recent work has also considered the costs associated with
ML-systems~\cite{sculley2015hidden, breck2017ml, bronzino2019inferring}.
Bronzino \etal{} addressed the problem of inferring the quality of video
streaming applications from encrypted traffic and classified the possible set of
features based on their corresponding layer in the network stack. This
categorization enabled the authors to logically reason about the cost associated
with each features sets. The observations about the tradeoffs between model
accuracy and systems costs for a specific problem motivated us to explore this
problem in general.
Sculley \etal{}, \cite{sculley2015hidden}
and Breck \etal{}~\cite{breck2017ml} investigated the hidden ``technical debt''
that incurs during the development and deployment of ML systems; 
the authors discuss system-level factors that increase the
maintenance costs of real-world ML systems over time (\eg,~unstable or
underutilized data, dependencies on proprietary packages, entanglement of input
signals, to name a few). \sysname{} builds on this work,
developing techniques to explore and mitigate technical debt associated with
data representation.

\section{Conclusion}

This paper introduces \system{}, which permits consideration of both model
accuracy and the systems-related costs of machine learning models trained on
network traffic representations to make predictions concerning performance and
security.  We show the need for exploring more flexible representations first
by showing that today's default representations result in lower model
accuracy.  We present the design and implementation of \system{} and apply it
to two use case studies: video quality performance inference and malware
detection.  This work has demonstrated both the need and the {\em potential}
for exploring how different data representations can affect model accuracy,
laying the groundwork for future work along multiple avenues, including
automated exploration of data representations, systems-level optimizations to
improve traffic processing capabilities and rates, and follow-up work that
considers the design of processing hardware in concert with the need for
specific data representations that result in high model accuracy across a
range of inference problems.  To enable the community to explore these
benefits on a wider range of problems, we have both released \system{} as
open-source software, as well as the evaluation in this paper.
\label{p:lastpage}
\end{sloppypar}

\bibliographystyle{ACM-Reference-Format}
\bibliography{bib}


\begin{thebibliography}{51}


\ifx \showCODEN    \undefined \def \showCODEN     #1{\unskip}     \fi
\ifx \showDOI      \undefined \def \showDOI       #1{#1}\fi
\ifx \showISBNx    \undefined \def \showISBNx     #1{\unskip}     \fi
\ifx \showISBNxiii \undefined \def \showISBNxiii  #1{\unskip}     \fi
\ifx \showISSN     \undefined \def \showISSN      #1{\unskip}     \fi
\ifx \showLCCN     \undefined \def \showLCCN      #1{\unskip}     \fi
\ifx \shownote     \undefined \def \shownote      #1{#1}          \fi
\ifx \showarticletitle \undefined \def \showarticletitle #1{#1}   \fi
\ifx \showURL      \undefined \def \showURL       {\relax}        \fi
\providecommand\bibfield[2]{#2}
\providecommand\bibinfo[2]{#2}
\providecommand\natexlab[1]{#1}
\providecommand\showeprint[2][]{arXiv:#2}

\bibitem[\protect\citeauthoryear{Borders, Springer, and Burnside}{Borders
  et~al\mbox{.}}{2012}]%
        {180224}
\bibfield{author}{\bibinfo{person}{Kevin Borders}, \bibinfo{person}{Jonathan
  Springer}, {and} \bibinfo{person}{Matthew Burnside}.}
  \bibinfo{year}{2012}\natexlab{}.
\newblock \showarticletitle{Chimera: A Declarative Language for Streaming
  Network Traffic Analysis}. In \bibinfo{booktitle}{{\em Presented as part of
  the 21st {USENIX} Security Symposium ({USENIX} Security 12)}}.
  \bibinfo{publisher}{{USENIX}}, \bibinfo{address}{Bellevue, WA},
  \bibinfo{pages}{365--379}.
\newblock
\showISBNx{978-931971-95-9}
\showURL{%
\url{https://www.usenix.org/conference/usenixsecurity12/technical-sessions/presentation/borders}}


\bibitem[\protect\citeauthoryear{Borgolte, Chattopadhyay, Feamster, Kshirsagar,
  Holland, Hounsel, and Schmitt}{Borgolte et~al\mbox{.}}{2019}]%
        {borgolte2019dns}
\bibfield{author}{\bibinfo{person}{Kevin Borgolte}, \bibinfo{person}{Tithi
  Chattopadhyay}, \bibinfo{person}{Nick Feamster}, \bibinfo{person}{Mihir
  Kshirsagar}, \bibinfo{person}{Jordan Holland}, \bibinfo{person}{Austin
  Hounsel}, {and} \bibinfo{person}{Paul Schmitt}.}
  \bibinfo{year}{2019}\natexlab{}.
\newblock \showarticletitle{How DNS over HTTPS is Reshaping Privacy,
  Performance, and Policy in the Internet Ecosystem}.
\newblock \bibinfo{journal}{{\em Performance, and Policy in the Internet
  Ecosystem (July 27, 2019)\/}} (\bibinfo{year}{2019}).
\newblock


\bibitem[\protect\citeauthoryear{Boutaba, Salahuddin, Limam, Ayoubi, Shahriar,
  Estrada-Solano, and Caicedo}{Boutaba et~al\mbox{.}}{2018}]%
        {boutaba2018comprehensive}
\bibfield{author}{\bibinfo{person}{Raouf Boutaba}, \bibinfo{person}{Mohammad~A
  Salahuddin}, \bibinfo{person}{Noura Limam}, \bibinfo{person}{Sara Ayoubi},
  \bibinfo{person}{Nashid Shahriar}, \bibinfo{person}{Felipe Estrada-Solano},
  {and} \bibinfo{person}{Oscar~M Caicedo}.} \bibinfo{year}{2018}\natexlab{}.
\newblock \showarticletitle{A comprehensive survey on machine learning for
  networking: evolution, applications and research opportunities}.
\newblock \bibinfo{journal}{{\em Journal of Internet Services and
  Applications\/}} \bibinfo{volume}{9}, \bibinfo{number}{1}
  (\bibinfo{year}{2018}), \bibinfo{pages}{16}.
\newblock


\bibitem[\protect\citeauthoryear{Breck, Cai, Nielsen, Salib, and Sculley}{Breck
  et~al\mbox{.}}{2017}]%
        {breck2017ml}
\bibfield{author}{\bibinfo{person}{Eric Breck}, \bibinfo{person}{Shanqing Cai},
  \bibinfo{person}{Eric Nielsen}, \bibinfo{person}{Michael Salib}, {and}
  \bibinfo{person}{D Sculley}.} \bibinfo{year}{2017}\natexlab{}.
\newblock \showarticletitle{The ml test score: A rubric for ml production
  readiness and technical debt reduction}. In \bibinfo{booktitle}{{\em 2017
  IEEE International Conference on Big Data (Big Data)}}. IEEE,
  \bibinfo{pages}{1123--1132}.
\newblock


\bibitem[\protect\citeauthoryear{Bronzino, Schmitt, Ayoubi, Martins, Teixeira,
  and Feamster}{Bronzino et~al\mbox{.}}{2019}]%
        {bronzino2019inferring}
\bibfield{author}{\bibinfo{person}{Francesco Bronzino}, \bibinfo{person}{Paul
  Schmitt}, \bibinfo{person}{Sara Ayoubi}, \bibinfo{person}{Guilherme Martins},
  \bibinfo{person}{Renata Teixeira}, {and} \bibinfo{person}{Nick Feamster}.}
  \bibinfo{year}{2019}\natexlab{}.
\newblock \showarticletitle{Inferring Streaming Video Quality from Encrypted
  Traffic: Practical Models and Deployment Experience}.
\newblock \bibinfo{journal}{{\em Proceedings of the ACM on Measurement and
  Analysis of Computing Systems\/}} \bibinfo{volume}{3}, \bibinfo{number}{3}
  (\bibinfo{year}{2019}), \bibinfo{pages}{1--25}.
\newblock


\bibitem[\protect\citeauthoryear{Claise}{Claise}{2004}]%
        {claise2004cisco}
\bibfield{author}{\bibinfo{person}{Benoit Claise}.}
  \bibinfo{year}{2004}\natexlab{}.
\newblock \bibinfo{booktitle}{{\em Cisco systems netflow services export
  version 9}}.
\newblock \bibinfo{type}{{T}echnical {R}eport}.
\newblock


\bibitem[\protect\citeauthoryear{Claise, Trammell, and Aitken.}{Claise
  et~al\mbox{.}}{2013}]%
        {ipfix}
\bibfield{author}{\bibinfo{person}{Benoit Claise}, \bibinfo{person}{Brian
  Trammell}, {and} \bibinfo{person}{Paul Aitken.}}
  \bibinfo{year}{2013}\natexlab{}.
\newblock \bibinfo{title}{Specification of the IP flow information export
  (IPFIX) protocol for the exchange of flow information}.
\newblock   (\bibinfo{year}{2013}).
\newblock
\newblock
\shownote{RFC 7011.}


\bibitem[\protect\citeauthoryear{corelight}{corelight}{2019}]%
        {corelight}
corelight \bibinfo{year}{2019}\natexlab{}.
\newblock \bibinfo{title}{{Corelight}}.
\newblock \bibinfo{howpublished}{\url{https://corelight.com/}}.
  (\bibinfo{year}{2019}).
\newblock


\bibitem[\protect\citeauthoryear{Cranor, Johnson, Spataschek, and
  Shkapenyuk}{Cranor et~al\mbox{.}}{2003}]%
        {cranor2003gigascope}
\bibfield{author}{\bibinfo{person}{Chuck Cranor}, \bibinfo{person}{Theodore
  Johnson}, \bibinfo{person}{Oliver Spataschek}, {and}
  \bibinfo{person}{Vladislav Shkapenyuk}.} \bibinfo{year}{2003}\natexlab{}.
\newblock \showarticletitle{Gigascope: a stream database for network
  applications}. In \bibinfo{booktitle}{{\em Proceedings of the 2003 ACM SIGMOD
  international conference on Management of data}}. ACM,
  \bibinfo{pages}{647--651}.
\newblock


\bibitem[\protect\citeauthoryear{deepfield}{deepfield}{2019}]%
        {deepfield}
deepfield \bibinfo{year}{2019}\natexlab{}.
\newblock \bibinfo{title}{{Deepfield}}.
\newblock \bibinfo{howpublished}{\url{https://deepfield.com/}}.
  (\bibinfo{year}{2019}).
\newblock


\bibitem[\protect\citeauthoryear{Deri et~al\mbox{.}}{Deri
  et~al\mbox{.}}{2004}]%
        {deri2004improving}
\bibfield{author}{\bibinfo{person}{Luca Deri} {et~al\mbox{.}}}
  \bibinfo{year}{2004}\natexlab{}.
\newblock \showarticletitle{Improving passive packet capture: Beyond device
  polling}. In \bibinfo{booktitle}{{\em Proceedings of SANE}},
  Vol.~\bibinfo{volume}{2004}. Amsterdam, Netherlands, \bibinfo{pages}{85--93}.
\newblock


\bibitem[\protect\citeauthoryear{dpdk}{dpdk}{2018}]%
        {dpdk}
dpdk \bibinfo{year}{2018}\natexlab{}.
\newblock \bibinfo{title}{{DPDK, Data Plane Development Kit}}.
\newblock \bibinfo{howpublished}{\url{https://www.dpdk.org/}}.
  (\bibinfo{year}{2018}).
\newblock


\bibitem[\protect\citeauthoryear{Duce}{Duce}{2003}]%
        {png}
\bibfield{author}{\bibinfo{person}{David Duce}.}
  \bibinfo{year}{2003}\natexlab{}.
\newblock \bibinfo{title}{Portable Network Graphics (PNG) Specification (Second
  Edition)}.
\newblock   (\bibinfo{year}{2003}).
\newblock
\newblock
\shownote{W3C Recommendation.}


\bibitem[\protect\citeauthoryear{Estan and Varghese}{Estan and
  Varghese}{2002}]%
        {Estan:2002:NDT:633025.633056}
\bibfield{author}{\bibinfo{person}{Cristian Estan} {and}
  \bibinfo{person}{George Varghese}.} \bibinfo{year}{2002}\natexlab{}.
\newblock \showarticletitle{New Directions in Traffic Measurement and
  Accounting}. In \bibinfo{booktitle}{{\em Proceedings of the 2002 Conference
  on Applications, Technologies, Architectures, and Protocols for Computer
  Communications}} {\em (\bibinfo{series}{SIGCOMM '02})}.
  \bibinfo{publisher}{ACM}, \bibinfo{address}{New York, NY, USA},
  \bibinfo{pages}{323--336}.
\newblock
\showISBNx{1-58113-570-X}
\showDOI{%
\url{https://doi.org/10.1145/633025.633056}}


\bibitem[\protect\citeauthoryear{Finamore, Mellia, Meo, Munaf{\`o}, and
  Rossi}{Finamore et~al\mbox{.}}{2010}]%
        {finamore2010live}
\bibfield{author}{\bibinfo{person}{Alessandro Finamore}, \bibinfo{person}{Marco
  Mellia}, \bibinfo{person}{Michela Meo}, \bibinfo{person}{Maurizio~M
  Munaf{\`o}}, {and} \bibinfo{person}{Dario Rossi}.}
  \bibinfo{year}{2010}\natexlab{}.
\newblock \showarticletitle{Live traffic monitoring with tstat: Capabilities
  and experiences}. In \bibinfo{booktitle}{{\em International Conference on
  Wired/Wireless Internet Communications}}. Springer,
  \bibinfo{pages}{290--301}.
\newblock


\bibitem[\protect\citeauthoryear{golang}{golang}{2020}]%
        {golang}
golang \bibinfo{year}{2020}\natexlab{}.
\newblock \bibinfo{title}{{Go language}}.
\newblock \bibinfo{howpublished}{\url{https://golang.org/}}.
  (\bibinfo{year}{2020}).
\newblock


\bibitem[\protect\citeauthoryear{gopacket}{gopacket}{2020}]%
        {gopacket}
gopacket \bibinfo{year}{2020}\natexlab{}.
\newblock \bibinfo{title}{Go Packet Library}.
\newblock   (\bibinfo{year}{2020}).
\newblock
\newblock
\shownote{{\url{https://godoc.org/github.com/google/gopacket}}.}


\bibitem[\protect\citeauthoryear{Gupta, Harrison, Canini, Feamster, Rexford,
  and Willinger}{Gupta et~al\mbox{.}}{2018}]%
        {Gupta:2018:SQS:3230543.3230555}
\bibfield{author}{\bibinfo{person}{Arpit Gupta}, \bibinfo{person}{Rob
  Harrison}, \bibinfo{person}{Marco Canini}, \bibinfo{person}{Nick Feamster},
  \bibinfo{person}{Jennifer Rexford}, {and} \bibinfo{person}{Walter
  Willinger}.} \bibinfo{year}{2018}\natexlab{}.
\newblock \showarticletitle{Sonata: Query-driven Streaming Network Telemetry}.
  In \bibinfo{booktitle}{{\em Proceedings of the 2018 Conference of the ACM
  Special Interest Group on Data Communication}} {\em (\bibinfo{series}{SIGCOMM
  '18})}. \bibinfo{publisher}{ACM}, \bibinfo{address}{New York, NY, USA},
  \bibinfo{pages}{357--371}.
\newblock
\showISBNx{978-1-4503-5567-4}
\showDOI{%
\url{https://doi.org/10.1145/3230543.3230555}}


\bibitem[\protect\citeauthoryear{Gutterman, Guo, Arora, Wang, Wu, Katz-Bassett,
  and Zussman}{Gutterman et~al\mbox{.}}{2019}]%
        {gutterman2019requet}
\bibfield{author}{\bibinfo{person}{Craig Gutterman}, \bibinfo{person}{Katherine
  Guo}, \bibinfo{person}{Sarthak Arora}, \bibinfo{person}{Xiaoyang Wang},
  \bibinfo{person}{Les Wu}, \bibinfo{person}{Ethan Katz-Bassett}, {and}
  \bibinfo{person}{Gil Zussman}.} \bibinfo{year}{2019}\natexlab{}.
\newblock \showarticletitle{Requet: Real-time qoe detection for encrypted
  youtube traffic}. In \bibinfo{booktitle}{{\em Proceedings of the 10th ACM
  Multimedia Systems Conference}}. \bibinfo{pages}{48--59}.
\newblock


\bibitem[\protect\citeauthoryear{Holland, Schmitt, Feamster, and
  Mittal}{Holland et~al\mbox{.}}{2020}]%
        {DBLP:journals/corr/abs-2008-02695}
\bibfield{author}{\bibinfo{person}{Jordan Holland}, \bibinfo{person}{Paul
  Schmitt}, \bibinfo{person}{Nick Feamster}, {and} \bibinfo{person}{Prateek
  Mittal}.} \bibinfo{year}{2020}\natexlab{}.
\newblock \showarticletitle{nPrint: {A} Standard Data Representation for
  Network Traffic Analysis}.
\newblock  (\bibinfo{year}{2020}).
\newblock
\showeprint[arxiv]{2008.02695}
\showURL{%
\url{https://arxiv.org/abs/2008.02695}}


\bibitem[\protect\citeauthoryear{kentik}{kentik}{2019}]%
        {kentik}
kentik \bibinfo{year}{2019}\natexlab{}.
\newblock \bibinfo{title}{{Kentik}}.
\newblock \bibinfo{howpublished}{\url{https://kentik.com/}}.
  (\bibinfo{year}{2019}).
\newblock


\bibitem[\protect\citeauthoryear{Krishnamoorthi, Carlsson, Halepovic, and
  Petajan}{Krishnamoorthi et~al\mbox{.}}{2017}]%
        {krishnamoorthi2017buffest}
\bibfield{author}{\bibinfo{person}{Vengatanathan Krishnamoorthi},
  \bibinfo{person}{Niklas Carlsson}, \bibinfo{person}{Emir Halepovic}, {and}
  \bibinfo{person}{Eric Petajan}.} \bibinfo{year}{2017}\natexlab{}.
\newblock \showarticletitle{{BUFFEST}: Predicting Buffer Conditions and
  Real-time Requirements of {HTTP} (S) Adaptive Streaming Clients}. In
  \bibinfo{booktitle}{{\em Proceedings of the 8th ACM on Multimedia Systems
  Conference}}. ACM, \bibinfo{pages}{76--87}.
\newblock


\bibitem[\protect\citeauthoryear{Kumar, Sung, Xu, and Wang}{Kumar
  et~al\mbox{.}}{2004}]%
        {kumar2004data}
\bibfield{author}{\bibinfo{person}{Abhishek Kumar}, \bibinfo{person}{Minho
  Sung}, \bibinfo{person}{Jun~Jim Xu}, {and} \bibinfo{person}{Jia Wang}.}
  \bibinfo{year}{2004}\natexlab{}.
\newblock \showarticletitle{Data streaming algorithms for efficient and
  accurate estimation of flow size distribution}. In \bibinfo{booktitle}{{\em
  ACM SIGMETRICS Performance Evaluation Review}}, Vol.~\bibinfo{volume}{32}.
  ACM, \bibinfo{pages}{177--188}.
\newblock


\bibitem[\protect\citeauthoryear{Liu, Manousis, Vorsanger, Sekar, and
  Braverman}{Liu et~al\mbox{.}}{2016}]%
        {liu2016one}
\bibfield{author}{\bibinfo{person}{Zaoxing Liu}, \bibinfo{person}{Antonis
  Manousis}, \bibinfo{person}{Gregory Vorsanger}, \bibinfo{person}{Vyas Sekar},
  {and} \bibinfo{person}{Vladimir Braverman}.} \bibinfo{year}{2016}\natexlab{}.
\newblock \showarticletitle{One sketch to rule them all: Rethinking network
  flow monitoring with univmon}. In \bibinfo{booktitle}{{\em Proceedings of the
  2016 ACM SIGCOMM Conference}}. ACM, \bibinfo{pages}{101--114}.
\newblock


\bibitem[\protect\citeauthoryear{Mangla, Halepovic, Ammar, and Zegura}{Mangla
  et~al\mbox{.}}{2019}]%
        {mangla2019using}
\bibfield{author}{\bibinfo{person}{Tarun Mangla}, \bibinfo{person}{Emir
  Halepovic}, \bibinfo{person}{Mostafa Ammar}, {and} \bibinfo{person}{Ellen
  Zegura}.} \bibinfo{year}{2019}\natexlab{}.
\newblock \showarticletitle{Using session modeling to estimate HTTP-based video
  QoE metrics from encrypted network traffic}.
\newblock \bibinfo{journal}{{\em IEEE Transactions on Network and Service
  Management\/}} \bibinfo{volume}{16}, \bibinfo{number}{3}
  (\bibinfo{year}{2019}), \bibinfo{pages}{1086--1099}.
\newblock


\bibitem[\protect\citeauthoryear{Mar{\'\i}n, Casas, and Capdehourat}{Mar{\'\i}n
  et~al\mbox{.}}{2018}]%
        {marin2018rawpower}
\bibfield{author}{\bibinfo{person}{Gonzalo Mar{\'\i}n}, \bibinfo{person}{Pedro
  Casas}, {and} \bibinfo{person}{Germ{\'a}n Capdehourat}.}
  \bibinfo{year}{2018}\natexlab{}.
\newblock \showarticletitle{Rawpower: Deep learning based anomaly detection
  from raw network traffic measurements}. In \bibinfo{booktitle}{{\em
  Proceedings of the ACM SIGCOMM 2018 Conference on Posters and Demos}}.
  \bibinfo{pages}{75--77}.
\newblock


\bibitem[\protect\citeauthoryear{Mar{\'\i}n, Casas, and Capdehourat}{Mar{\'\i}n
  et~al\mbox{.}}{2020}]%
        {marin2020deepmal}
\bibfield{author}{\bibinfo{person}{Gonzalo Mar{\'\i}n}, \bibinfo{person}{Pedro
  Casas}, {and} \bibinfo{person}{Germ{\'a}n Capdehourat}.}
  \bibinfo{year}{2020}\natexlab{}.
\newblock \showarticletitle{DeepMAL--Deep Learning Models for Malware Traffic
  Detection and Classification}.
\newblock \bibinfo{journal}{{\em arXiv preprint arXiv:2003.04079\/}}
  (\bibinfo{year}{2020}).
\newblock


\bibitem[\protect\citeauthoryear{Mazhar and Shafiq}{Mazhar and Shafiq}{2018a}]%
        {mazhar2018real}
\bibfield{author}{\bibinfo{person}{M~Hammad Mazhar} {and}
  \bibinfo{person}{Zubair Shafiq}.} \bibinfo{year}{2018}\natexlab{a}.
\newblock \showarticletitle{Real-time video quality of experience monitoring
  for https and quic}. In \bibinfo{booktitle}{{\em IEEE INFOCOM 2018-IEEE
  Conference on Computer Communications}}. IEEE, \bibinfo{pages}{1331--1339}.
\newblock


\bibitem[\protect\citeauthoryear{Mazhar and Shafiq}{Mazhar and Shafiq}{2018b}]%
        {infocomPaper}
\bibfield{author}{\bibinfo{person}{M.~Hammad Mazhar} {and}
  \bibinfo{person}{Zubair Shafiq}.} \bibinfo{year}{2018}\natexlab{b}.
\newblock \showarticletitle{Real-time Video Quality of Experience Monitoring
  for {HTTPS and QUIC}}. In \bibinfo{booktitle}{{\em INFOCOM, 2018 Proceedings
  IEEE}}. IEEE.
\newblock


\bibitem[\protect\citeauthoryear{Mellia, Carpani, and Cigno}{Mellia
  et~al\mbox{.}}{2003}]%
        {mellia2003tstat}
\bibfield{author}{\bibinfo{person}{Marco Mellia}, \bibinfo{person}{Andrea
  Carpani}, {and} \bibinfo{person}{Renato~Lo Cigno}.}
  \bibinfo{year}{2003}\natexlab{}.
\newblock \showarticletitle{Tstat: TCP statistic and analysis tool}. In
  \bibinfo{booktitle}{{\em International Workshop on Quality of Service in
  Multiservice IP Networks}}. Springer, \bibinfo{pages}{145--157}.
\newblock


\bibitem[\protect\citeauthoryear{Narayana, Sivaraman, Nathan, Goyal, Arun,
  Alizadeh, Jeyakumar, and Kim}{Narayana et~al\mbox{.}}{2017}]%
        {narayana2017language}
\bibfield{author}{\bibinfo{person}{Srinivas Narayana}, \bibinfo{person}{Anirudh
  Sivaraman}, \bibinfo{person}{Vikram Nathan}, \bibinfo{person}{Prateesh
  Goyal}, \bibinfo{person}{Venkat Arun}, \bibinfo{person}{Mohammad Alizadeh},
  \bibinfo{person}{Vimalkumar Jeyakumar}, {and} \bibinfo{person}{Changhoon
  Kim}.} \bibinfo{year}{2017}\natexlab{}.
\newblock \showarticletitle{Language-directed hardware design for network
  performance monitoring}. In \bibinfo{booktitle}{{\em Proceedings of the
  Conference of the ACM Special Interest Group on Data Communication}}. ACM,
  \bibinfo{pages}{85--98}.
\newblock


\bibitem[\protect\citeauthoryear{netvcr}{netvcr}{2020}]%
        {netvcr}
netvcr \bibinfo{year}{2020}\natexlab{}.
\newblock \bibinfo{title}{{NIKSUN NetVCR}}.
\newblock
  \bibinfo{howpublished}{\url{https://www.niksun.com/product.php?id=110}}.
  (\bibinfo{year}{2020}).
\newblock


\bibitem[\protect\citeauthoryear{Nguyen and Armitage}{Nguyen and
  Armitage}{2008}]%
        {nguyen2008survey}
\bibfield{author}{\bibinfo{person}{Thuy~TT Nguyen} {and}
  \bibinfo{person}{Grenville Armitage}.} \bibinfo{year}{2008}\natexlab{}.
\newblock \showarticletitle{A survey of techniques for internet traffic
  classification using machine learning}.
\newblock \bibinfo{journal}{{\em IEEE communications surveys \& tutorials\/}}
  \bibinfo{volume}{10}, \bibinfo{number}{4} (\bibinfo{year}{2008}),
  \bibinfo{pages}{56--76}.
\newblock


\bibitem[\protect\citeauthoryear{Orebaugh, Ramirez, and Beale}{Orebaugh
  et~al\mbox{.}}{2006}]%
        {orebaugh2006wireshark}
\bibfield{author}{\bibinfo{person}{Angela Orebaugh}, \bibinfo{person}{Gilbert
  Ramirez}, {and} \bibinfo{person}{Jay Beale}.}
  \bibinfo{year}{2006}\natexlab{}.
\newblock \bibinfo{booktitle}{{\em Wireshark \& Ethereal network protocol
  analyzer toolkit}}.
\newblock \bibinfo{publisher}{Elsevier}.
\newblock


\bibitem[\protect\citeauthoryear{Paxson}{Paxson}{1999}]%
        {paxson1999bro}
\bibfield{author}{\bibinfo{person}{Vern Paxson}.}
  \bibinfo{year}{1999}\natexlab{}.
\newblock \showarticletitle{Bro: a system for detecting network intruders in
  real-time}.
\newblock \bibinfo{journal}{{\em Computer networks\/}} \bibinfo{volume}{31},
  \bibinfo{number}{23-24} (\bibinfo{year}{1999}), \bibinfo{pages}{2435--2463}.
\newblock


\bibitem[\protect\citeauthoryear{Plonka and Barford}{Plonka and
  Barford}{2011}]%
        {plonka2011flexible}
\bibfield{author}{\bibinfo{person}{David Plonka} {and} \bibinfo{person}{Paul
  Barford}.} \bibinfo{year}{2011}\natexlab{}.
\newblock \showarticletitle{Flexible traffic and host profiling via DNS
  rendezvous}. In \bibinfo{booktitle}{{\em Workshop Satin}}.
\newblock


\bibitem[\protect\citeauthoryear{Reddy, Wing, and Patil}{Reddy
  et~al\mbox{.}}{2017}]%
        {reddy2017dns}
\bibfield{author}{\bibinfo{person}{Tirumaleswar Reddy}, \bibinfo{person}{Dan
  Wing}, {and} \bibinfo{person}{Prashanth Patil}.}
  \bibinfo{year}{2017}\natexlab{}.
\newblock \showarticletitle{Dns over datagram transport layer security (dtls)}.
\newblock \bibinfo{journal}{{\em RFC 8094\/}} (\bibinfo{year}{2017}).
\newblock


\bibitem[\protect\citeauthoryear{Roesch et~al\mbox{.}}{Roesch
  et~al\mbox{.}}{1999}]%
        {roesch1999snort}
\bibfield{author}{\bibinfo{person}{Martin Roesch} {et~al\mbox{.}}}
  \bibinfo{year}{1999}\natexlab{}.
\newblock \showarticletitle{Snort: Lightweight intrusion detection for
  networks.}. In \bibinfo{booktitle}{{\em Lisa}}, Vol.~\bibinfo{volume}{99}.
  \bibinfo{pages}{229--238}.
\newblock


\bibitem[\protect\citeauthoryear{Sculley, Holt, Golovin, Davydov, Phillips,
  Ebner, Chaudhary, Young, Crespo, and Dennison}{Sculley et~al\mbox{.}}{2015}]%
        {sculley2015hidden}
\bibfield{author}{\bibinfo{person}{David Sculley}, \bibinfo{person}{Gary Holt},
  \bibinfo{person}{Daniel Golovin}, \bibinfo{person}{Eugene Davydov},
  \bibinfo{person}{Todd Phillips}, \bibinfo{person}{Dietmar Ebner},
  \bibinfo{person}{Vinay Chaudhary}, \bibinfo{person}{Michael Young},
  \bibinfo{person}{Jean-Francois Crespo}, {and} \bibinfo{person}{Dan
  Dennison}.} \bibinfo{year}{2015}\natexlab{}.
\newblock \showarticletitle{Hidden technical debt in machine learning systems}.
  In \bibinfo{booktitle}{{\em Advances in neural information processing
  systems}}. \bibinfo{pages}{2503--2511}.
\newblock


\bibitem[\protect\citeauthoryear{Sharafaldin, Lashkari, and
  Ghorbani}{Sharafaldin et~al\mbox{.}}{2018}]%
        {sharafaldin2018toward}
\bibfield{author}{\bibinfo{person}{Iman Sharafaldin},
  \bibinfo{person}{Arash~Habibi Lashkari}, {and} \bibinfo{person}{Ali~A
  Ghorbani}.} \bibinfo{year}{2018}\natexlab{}.
\newblock \showarticletitle{Toward generating a new intrusion detection dataset
  and intrusion traffic characterization.}. In \bibinfo{booktitle}{{\em
  ICISSP}}. \bibinfo{pages}{108--116}.
\newblock


\bibitem[\protect\citeauthoryear{Singh and Nene}{Singh and Nene}{2013}]%
        {singh2013survey}
\bibfield{author}{\bibinfo{person}{Jayveer Singh} {and}
  \bibinfo{person}{Manisha~J Nene}.} \bibinfo{year}{2013}\natexlab{}.
\newblock \showarticletitle{A survey on machine learning techniques for
  intrusion detection systems}.
\newblock \bibinfo{journal}{{\em International Journal of Advanced Research in
  Computer and Communication Engineering\/}} \bibinfo{volume}{2},
  \bibinfo{number}{11} (\bibinfo{year}{2013}), \bibinfo{pages}{4349--4355}.
\newblock


\bibitem[\protect\citeauthoryear{tcpdump}{tcpdump}{2020}]%
        {pcap}
tcpdump \bibinfo{year}{2020}\natexlab{}.
\newblock \bibinfo{title}{tcpdump and libpcap}.
\newblock \bibinfo{howpublished}{\url{https://www.tcpdump.org/}}.
  (\bibinfo{year}{2020}).
\newblock


\bibitem[\protect\citeauthoryear{traffica}{traffica}{2020}]%
        {traffica}
traffica \bibinfo{year}{2020}\natexlab{}.
\newblock \bibinfo{title}{{Nokia Traffica}}.
\newblock
  \bibinfo{howpublished}{\url{https://www.nokia.com/networks/products/traffica/}}.
    (\bibinfo{year}{2020}).
\newblock


\bibitem[\protect\citeauthoryear{tshark}{tshark}{2020}]%
        {www-tshark}
tshark \bibinfo{year}{2020}\natexlab{}.
\newblock \bibinfo{title}{{Tshark: terminal-based Wireshark}}.
\newblock
  \bibinfo{howpublished}{\url{https://www.wireshark.org/docs/wsug_html_chunked/AppToolstshark.html}}.
    (\bibinfo{year}{2020}).
\newblock


\bibitem[\protect\citeauthoryear{Wang, Sheng, Wang, Zeng, Ye, Huang, and
  Zhu}{Wang et~al\mbox{.}}{2017a}]%
        {wang2017hast}
\bibfield{author}{\bibinfo{person}{Wei Wang}, \bibinfo{person}{Yiqiang Sheng},
  \bibinfo{person}{Jinlin Wang}, \bibinfo{person}{Xuewen Zeng},
  \bibinfo{person}{Xiaozhou Ye}, \bibinfo{person}{Yongzhong Huang}, {and}
  \bibinfo{person}{Ming Zhu}.} \bibinfo{year}{2017}\natexlab{a}.
\newblock \showarticletitle{HAST-IDS: Learning hierarchical spatial-temporal
  features using deep neural networks to improve intrusion detection}.
\newblock \bibinfo{journal}{{\em IEEE Access\/}}  \bibinfo{volume}{6}
  (\bibinfo{year}{2017}), \bibinfo{pages}{1792--1806}.
\newblock


\bibitem[\protect\citeauthoryear{Wang, Zhu, Zeng, Ye, and Sheng}{Wang
  et~al\mbox{.}}{2017b}]%
        {wang2017malware}
\bibfield{author}{\bibinfo{person}{Wei Wang}, \bibinfo{person}{Ming Zhu},
  \bibinfo{person}{Xuewen Zeng}, \bibinfo{person}{Xiaozhou Ye}, {and}
  \bibinfo{person}{Yiqiang Sheng}.} \bibinfo{year}{2017}\natexlab{b}.
\newblock \showarticletitle{Malware traffic classification using convolutional
  neural network for representation learning}. In \bibinfo{booktitle}{{\em 2017
  International Conference on Information Networking (ICOIN)}}. IEEE,
  \bibinfo{pages}{712--717}.
\newblock


\bibitem[\protect\citeauthoryear{Yang, Jiang, Liu, Huang, Gong, Zhou, Miao, Li,
  and Uhlig}{Yang et~al\mbox{.}}{2018}]%
        {yang2018elastic}
\bibfield{author}{\bibinfo{person}{Tong Yang}, \bibinfo{person}{Jie Jiang},
  \bibinfo{person}{Peng Liu}, \bibinfo{person}{Qun Huang},
  \bibinfo{person}{Junzhi Gong}, \bibinfo{person}{Yang Zhou},
  \bibinfo{person}{Rui Miao}, \bibinfo{person}{Xiaoming Li}, {and}
  \bibinfo{person}{Steve Uhlig}.} \bibinfo{year}{2018}\natexlab{}.
\newblock \showarticletitle{Elastic sketch: Adaptive and fast network-wide
  measurements}. In \bibinfo{booktitle}{{\em Proceedings of the 2018 Conference
  of the ACM Special Interest Group on Data Communication}}. ACM,
  \bibinfo{pages}{561--575}.
\newblock


\bibitem[\protect\citeauthoryear{Yu, Zhu, Arzani, Fonseca, Zhang, Deng, and
  Yuan}{Yu et~al\mbox{.}}{2019}]%
        {yu2019dshark}
\bibfield{author}{\bibinfo{person}{Da Yu}, \bibinfo{person}{Yibo Zhu},
  \bibinfo{person}{Behnaz Arzani}, \bibinfo{person}{Rodrigo Fonseca},
  \bibinfo{person}{Tianrong Zhang}, \bibinfo{person}{Karl Deng}, {and}
  \bibinfo{person}{Lihua Yuan}.} \bibinfo{year}{2019}\natexlab{}.
\newblock \showarticletitle{dShark: A general, easy to program and scalable
  framework for analyzing in-network packet traces}. In
  \bibinfo{booktitle}{{\em 16th USENIX Symposium on Networked Systems Design
  and Implementation (NSDI 19)}}. \bibinfo{pages}{207--220}.
\newblock


\bibitem[\protect\citeauthoryear{Yu, Jose, and Miao}{Yu et~al\mbox{.}}{2013}]%
        {yu2013software}
\bibfield{author}{\bibinfo{person}{Minlan Yu}, \bibinfo{person}{Lavanya Jose},
  {and} \bibinfo{person}{Rui Miao}.} \bibinfo{year}{2013}\natexlab{}.
\newblock \showarticletitle{Software Defined Traffic Measurement with
  OpenSketch}. In \bibinfo{booktitle}{{\em Presented as part of the 10th USENIX
  Symposium on Networked Systems Design and Implementation (NSDI 13)}}.
  \bibinfo{pages}{29--42}.
\newblock


\bibitem[\protect\citeauthoryear{Yuan, Lin, Mishra, Marwaha, Alur, and
  Loo}{Yuan et~al\mbox{.}}{2017}]%
        {Yuan:2017:QNM:3098822.3098830}
\bibfield{author}{\bibinfo{person}{Yifei Yuan}, \bibinfo{person}{Dong Lin},
  \bibinfo{person}{Ankit Mishra}, \bibinfo{person}{Sajal Marwaha},
  \bibinfo{person}{Rajeev Alur}, {and} \bibinfo{person}{Boon~Thau Loo}.}
  \bibinfo{year}{2017}\natexlab{}.
\newblock \showarticletitle{Quantitative Network Monitoring with {NetQRE}}. In
  \bibinfo{booktitle}{{\em Proceedings of the Conference of the ACM Special
  Interest Group on Data Communication}} {\em (\bibinfo{series}{SIGCOMM '17})}.
  \bibinfo{publisher}{ACM}, \bibinfo{address}{New York, NY, USA},
  \bibinfo{pages}{99--112}.
\newblock
\showISBNx{978-1-4503-4653-5}
\showDOI{%
\url{https://doi.org/10.1145/3098822.3098830}}


\bibitem[\protect\citeauthoryear{Zhu, Kang, Cao, Greenberg, Lu, Mahajan, Maltz,
  Yuan, Zhang, Zhao, et~al\mbox{.}}{Zhu et~al\mbox{.}}{2015}]%
        {zhu2015packet}
\bibfield{author}{\bibinfo{person}{Yibo Zhu}, \bibinfo{person}{Nanxi Kang},
  \bibinfo{person}{Jiaxin Cao}, \bibinfo{person}{Albert Greenberg},
  \bibinfo{person}{Guohan Lu}, \bibinfo{person}{Ratul Mahajan},
  \bibinfo{person}{Dave Maltz}, \bibinfo{person}{Lihua Yuan},
  \bibinfo{person}{Ming Zhang}, \bibinfo{person}{Ben~Y Zhao}, {et~al\mbox{.}}}
  \bibinfo{year}{2015}\natexlab{}.
\newblock \showarticletitle{Packet-level telemetry in large datacenter
  networks}. In \bibinfo{booktitle}{{\em ACM SIGCOMM Computer Communication
  Review}}, Vol.~\bibinfo{volume}{45}. ACM, \bibinfo{pages}{479--491}.
\newblock


\end{thebibliography}

\appendix
\section{Video Quality Inference Configuration Details}\label{app:video}

This section provides additional details regarding the configuration and
implementation in \sysname{} of the video quality inference use case.

\subsection{Configuration}
Listing~\ref{lst:netflix_config} shows a complete configuration used to collect
Netflix traffic features. The configuration includes filters for known Netflix
domains, as well Netflix owned IP network prefixes. Further, the configuration
instructs \system{} to collect the features for the three different classes
described in Section~\ref{sec:deployment}, \ie, PacketCounters, TCPCounters,
and VideoSegments. Finally, the statistics produced from these features are
collected at ten seconds intervals. 

\begin{figure}[h]
  \begin{lstlisting}[language=json, caption=Configuration to capture video features for Netflix., captionpos=b, label={lst:netflix_config}]
    {
      "Name": "Netflix",
      "Filter": {
        "DomainsString": ["netflix.com","nflxvideo.net","nflximg.net","nflxext.com","nflximg.com","nflxso.net"],
        "Prefixes": "23.246.0.0/18", "37.77.184.0/21", "45.57.0.0/17", "64.120.128.0/17", "66.197.128.0/17", "108.175.32.0/20", "185.2.220.0/22", "185.9.188.0/22", "192.173.64.0/18", "198.38.96.0/19", "198.45.48.0/20", "208.75.79.0/24", "2620:10c:7000::/44", "2a00:86c0::/32"]
      },
      "Collect": [PacketCounters, TCPCounters, VideoSegments],
      "Emit": 10
    }
  \end{lstlisting}
\end{figure}

\subsection{Implementation}
Listing~\ref{lst:video_counters} shows the \texttt{AddPacket} implementation
used to collect video segments information for the video quality inference use
case. The function implements the technique first presented by Vengatanathan
\etal~\cite{krishnamoorthi2017buffest} who showed how video segment
information can be extracted by observing patterns in upstream traffic. In
particular, this method uses upstream requests times to break down the
stream of downstream packets into video segments. 

\begin{figure}[h]
  \begin{lstlisting}[language=Go, caption=Implementing the VideoSegments counters, captionpos=b, label={lst:video_counters}]
  // VideoSegment is used to keep track of segments in 
  // download
  type VideoSegment struct {
    Len       int64
    Seq       int64
    TsStart   int64
    TsEnd     int64
    LastPkt   int64
    DownPkts  int64
    DonwBytes int64
    MaxDSeq   int64
  }

  // VideoSegments is the flow stats structure used 
  // to store segments information
  type VideoSegments struct {
    CompleteSegments  []VideoSegment
    RunningSegment    VideoSegment
  }

  const (
    // QUICHeaderLen represents the minimum length in
    // bytes to determine when a QUIC upstream packet 
    // contains payload
    QUICHeaderLen = 100
  )

  // AddPacket updates the flow states based on the 
  // packet pkt
  func (vf *VideoSegments) AddPacket(pkt *network.Packet) error {
    if pkt.Dir == network.TrafficOut {
      if (pkt.IsTCP && pkt.DataLength > 0) || (!pkt.IsTCP && pkt.DataLength > QUICHeaderLen) {
        if vf.RunningSegment.TsStart != 0 && vf.RunningSegment.DownPkts > 0 {
          vf.RunningSegment.TsEnd = vf.RunningSegment.LastPkt
          vf.CompleteSegments = append(vf.CompleteSegments, vf.RunningSegment)
        }
        vf.RunningSegment = VideoSegment{Len: int64(pkt.Length), TsStart: pkt.TStamp, Seq: int64(pkt.Tcp.Seq)}
      }
    } else if pkt.DataLength > 0 {
      vf.RunningSegment.DownPkts++
      vf.RunningSegment.DonwBytes += int64(pkt.DataLength)

      if int64(pkt.Tcp.Seq) > vf.RunningSegment.MaxDSeq {
        vf.RunningSegment.MaxDSeq = int64(pkt.Tcp.Seq)
      }
      if pkt.TStamp > vf.RunningSegment.TsEnd {
        vf.RunningSegment.LastPkt = pkt.TStamp
      }
    }
    return nil
  }
  \end{lstlisting}
\end{figure}

\label{p:totalpage}
\end{document}